\documentclass[12pt]{article}
\usepackage{a41}
\usepackage{cite}
\usepackage{amsmath}
\usepackage{amssymb}

\usepackage{url}
\usepackage{booktabs}                      
\usepackage{epsfig} %
\usepackage{float} %
\usepackage{verbatim} %
\usepackage{color} %
\usepackage{eufrak}
\usepackage[english]{babel}
\usepackage[latin1]{inputenc}
\usepackage[T1]{fontenc}
\usepackage{ae}

\usepackage[rflt]{floatflt}              
\usepackage{float}
\usepackage{slashed}
\def\b0{\beta_0}

\usepackage{array}

\newcommand{\BLCB}{\left\{ \phantom{\rule{0.1mm}{0.57cm}} \! \right.}
\newcommand{\BLB}{\left[ \phantom{\rule{0.1mm}{0.56cm}} \! \right.}
\newcommand{\BRB}{\left. \! \phantom{\rule{0.1mm}{0.56cm}} \right] }
\newcommand{\BRCB}{\left. \! \phantom{\rule{0.1mm}{0.56cm}} \right\}}

\usepackage[english]{babel}
\usepackage[latin1]{inputenc}
\usepackage[T1]{fontenc}
\usepackage{ae}

\usepackage{url}

\usepackage{amsmath, amsthm, amssymb}
\newtheorem{thm}{Theorem}[section]

\newtheorem{definition}[thm]{Definition}

 \newcommand{\GeV}{\mathrm{GeV}}
 \newcommand{\keV}{\mathrm{keV}}
 
\newcommand{\HT}{{\rm H}^*}

\newcommand{\ep}{\varepsilon}

\usepackage{rotating}

\usepackage{graphicx}
\newcounter{mmacnt}
\def\restartmma{\setcounter{mmacnt}{0}}
\restartmma \catcode`|=\active
\def|#1|{\mathrm{#1}}
\catcode`|=12
\newenvironment{mma}{
 \par\smallskip
 \catcode`|=\active
 \parskip=0pt\parindent=0pt % locally
 \small
 \def\In##1\\{%
\def\linebreak{\hfill\break\null\qquad}%
\refstepcounter{mmacnt}
\hangindent=2.5em\hangafter=0
\leavevmode
\llap{\tiny\sffamily n[\arabic{mmacnt}]:=\kern.5em}%
\mathversion{bold}\footnotesize$\displaystyle##1$\normalsize
\mathversion{normal}\par
 }%
 \def\Print##1\\{%
\def\linebreak{\hfill\break}%
\hangindent=2.5em\hangafter=0
\leavevmode ##1\par}%
 \def\Out##1\\{%
\def\linebreak{$\hfill\break\null\hfill$}%
\kern\abovedisplayskip\par
\hangindent=2.5em\hangafter=0
\leavevmode
\llap{\tiny\sffamily Out[\arabic{mmacnt}]=\kern.5em}
\footnotesize$\displaystyle##1$\normalsize\hfill\null\par
\kern\belowdisplayskip
 }%
 \def\Warning##1##2\\{%
\def\linebreak{\hfill\break}%
\hangindent=2.5em\hangafter=0
\leavevmode
{\scriptsize##1 : ##2}\par}%
}{%
 \par\smallskip
}

\usepackage{color}

\newenvironment{fshaded}{%
\MakeFramed {\FrameRestore}
}%
{\endMakeFramed}

\def\b0{\beta_0}

\def\Gp0{{\Gamma^{'}_0}}
\def\Gp1{{\Gamma^{'}_1}}
\def\Gp2{{\Gamma^{'}_2}}

\allowdisplaybreaks[4]

\begin{document}
\setlength{\baselineskip}{0.515cm}

\sloppy
\thispagestyle{empty}
\begin{flushleft}
DESY 22--024
%  \hfill {\tt arXiv:2202.xxxxx[hep-ph]}
\\
DO-TH 22/04\\
TTP 22--011\\
SAGEX 22--19\\
\end{flushleft}

\mbox{}
\vspace*{\fill}
\begin{center}

{\LARGE\bf High Precision QED Initial State Corrections for}

\vspace*{2mm}
{\LARGE\bf \boldmath 
 $e^+ e^- \rightarrow \gamma^*/Z^*$ Annihilation}

\vspace{3cm} \large {\large J.~Bl\"umlein$^a$ and K.~Sch\"onwald$^{a,b}$ }

\vspace{1.cm}
\normalsize
{\it   $^a$~Deutsches Elektronen--Synchrotron DESY,}\\
{\it   Platanenallee 6, 15738 Zeuthen, Germany}

\vspace*{2mm}
{\it  $^b$~Institut f\"ur Theoretische Teilchenphysik,\\
Karlsruher Institut f\"ur Technologie (KIT), D-76128 Karlsruhe, Germany}

%%\today

\end{center}
\normalsize
\vspace{\fill}
\begin{abstract}
\noindent
The precise knowledge of the QED initial state corrections is of instrumental importance in studying 
high luminosity measurements in $e^+e^-$ annihilation at facilities like LEP, the International Linear Collider 
ILC, CLIC, a Giga-$Z$ facility, and the planned FCC\_ee. Logarithmic corrections of up to $O(\alpha^6 L^5)$ are
necessary with various subleading terms taken into account. This applies to both the inclusive measurement
of processes like $e^+ e^- \rightarrow \gamma^*/Z^*$ and also the forward--backward asymmetry. As has been shown 
recently, techniques from massive QCD, such as the computation of massive on--shell operator matrix elements, can 
be used for these calculations. We give an introduction to this topic and present both the calculation
methods and the numerical corrections having been reached so far.
\end{abstract}

\vspace*{\fill}
\noindent
% \numberwithin{equation}{section}
%%%%%%%%%%%%%%%%%%%%%%%%%%%%%%%%%%%%%%%%%%%%%%%%%%%%%%%%%%%%%%%%%%%%%%%%%%%%%%%%%%%%%%%%%%%%%%%%%%%%%%%%%%%%%%%%%%%%%%%%%%%%%%%%%%%
\newpage 

%--------------------------------------------------------------------------------------------------------
\section{Introduction}
\label{sec:1}
%--------------------------------------------------------------------------------------------------------

\vspace*{1mm}
\noindent
The precision physics for the process $e^+ e^- \rightarrow \gamma^*/Z^*$ started at LEP-1 in the early 1990ies
leading to a highly precise measurement of the $Z$--peak \cite{ALEPH:2005ab}. Among other radiative 
corrections the QED initial state corrections are of high importance in these measurements, since they
change the shape--parameters of the $Z$--peak significantly. The most far--reaching calculation
of the initial state QED corrections (ISR) of $O(\alpha^2)$, with $\alpha = 4 \pi a$ the fine structure 
constant, had been performed
in Ref.~\cite{Berends:1987ab} at that time, which had been a theoretical challenge and milestone. This 
calculation has never
been checked until 2011, some time after LEP had already been ended, when the same result was derived using massive
operator matrix elements (OMEs) in Ref.~\cite{Blumlein:2011mi}. Massive operator elements were known to present all
contributions except power corrections in QCD in the cases with external massless partons \cite{Buza:1995ie}. 
Therefore, one could try to apply this method also in the case of massive fermions, the electrons.
However, the present case is one with also massive external lines. After correcting typographical 
errors in \cite{Berends:1987ab} all logarithmic contributions of both calculations 
\cite{Berends:1987ab,Blumlein:2011mi} and the constant term at $O(\alpha)$ agreed, but not the result for the 
constant term of $O(\alpha^2)$. Furthermore, in one of the subprocesses Ref.~\cite{Berends:1987ab} agreed
with \cite{Kniehl:1988id}. Due to this, one option to explain the observed differences has been 
that the method of massive OMEs does not work in the case of massive external lines, or, otherwise,
that the results in Refs.~\cite{Berends:1987ab,Kniehl:1988id} need corrections.

{One way to unambiguously resolve these discrepancies} has been to repeat the complete calculation 
to $O(\alpha^2)$ without neglecting any term at intermediate steps and carry out the limit $m_e^2/s \rightarrow 0$ 
for the power corrections only in the final results. 
Such a calculation has not been fully possible in 2011, but became accessible in 2019,
again working on complete QCD calculation in which specific phase space integrals occurred. The organization
of these integrals as iterated integrals over various root--valued letters, cf. e.g.~\cite{Blumlein:2019qze,
Blumlein:2019zux}, and their reduction using shuffle algebras \cite{Blumlein:2003gb}, brought the solution to also 
organize the phase--space integrals in the case of QED. In this way, we firstly found the correct results 
for the fermionic non--singlet corrections of \cite{Berends:1987ab,Kniehl:1988id}, if the radiated particles
are electrons, but not muons, \cite{Blumlein:2019srk}. While our result agreed with 
\cite{Berends:1987ab,Kniehl:1988id} for muons, it disagreed for electron radiation, since there electron mass
terms had been neglected too early in Refs.~\cite{Berends:1987ab,Kniehl:1988id}. 
For Ref.~\cite{Berends:1987ab} this 
also applied to the pure singlet channel, which has been calculated earlier in the massless case in QCD in 
Ref.~\cite{SCHELLEKENS}, and been used as such in \cite{Berends:1987ab}. However, also here electron mass terms
cannot be neglected. We also newly calculated the bosonic contributions in 
\cite{Blumlein:2020jrf,Blumlein:2019pqb}, and fully confirmed the results given in Ref.~\cite{Blumlein:2011mi} by
the phase space integration method intended to be used in Ref.~\cite{Berends:1987ab}. Through this we proofed 
that the method of massive OMEs can also be used in the case of external massive fermionic lines.

{The technique of massive OMEs now also allows to access logarithmically enhanced 
  terms through the renormalization group. 
  With the known pieces it is possible to obtain}
three consecutive logarithmic orders in all higher 
orders in $\alpha$, as e.g.
$O(\alpha^3 L^3), O(\alpha^3 L^2), O(\alpha^3 L)$, etc., with $L = \ln(s/m_e^2)$ and $s$ the $cms$ energy.
We applied this technique to all corrections up to $O(\alpha^6 L^5)$ in Ref.~\cite{Ablinger:2020qvo} and to higher order corrections 
to the forward--backward asymmetry \cite{Blumlein:2021jdl}, although in this case new contributions have to be calculated. 
We also note that the universal QED corrections 
$O(\alpha^k L^k)$ are known to at least fifth order $k = 5$, 
cf.~\cite{Skrzypek:1992vk,Jezabek:1992bx,Przybycien:1992qe,
Blumlein:1996yz,Arbuzov:1999cq,Arbuzov:1999uq,Blumlein:2004bs,Blumlein:2007kx}.
Highly precise QED radiative corrections are of importance for the analysis of the precision measurements at LEP,
\cite{ALEPH:2005ab}, the planned international linear collider, ILC, \cite{ILC,Aihara:2019gcq,Mnich:2019}
CLIC \cite{CLIC}, the FCC\_ee \cite{FCCEE} and muon colliders \cite{Delahaye:2019omf}.

The paper is organized as follows. In Section~\ref{sec:2} we describe the direct calculation of the 
$O(\alpha^2)$ corrections due to initial state radiation and the factorized approach by using massive 
OMEs, which lead to the same results. For the $O(\alpha^2)$ ISR corrections we present numerical results
on the $Z^0$-peak and width, for the process $e^+e^- \rightarrow Z^0 H$ and $t\bar{t}$ production in the threshold 
region. The method of massive OMEs is then extended to higher orders
for the leading and two subleading logarithmic terms in Section~\ref{sec:3}. In Section~\ref{sec:4} we present 
precision results for the forward--backward asymmetry and Section~\ref{sec:5} contains the conclusions.
We also present numerical results on the calculated radiative corrections for different key observables. 
%--------------------------------------------------------------------------------------------------------
\section{Two--loop Corrections}
\label{sec:2}
%--------------------------------------------------------------------------------------------------------

\vspace*{1mm}
\noindent
After observing the differences in the results of Refs.~\cite{Blumlein:2011mi} and \cite{Berends:1987ab},
the only way to clarify this consisted in a direct calculation of the complete $O(a^2)$ initial state
corrections to $e^+ e^- \rightarrow \gamma^*/Z^*$ annihilation. Its result confirmed the results of 
Ref.~\cite{Blumlein:2011mi}. Due to this the explicit high energy factorization could be shown to $O(a^2)$
also for massive external fermions. Assuming this factorization to even higher orders, allows to calculate
the first three logarithmic correction terms to even higher order. 
These results are reported in Section~\ref{sec:3}. In the following 
we first summarize the complete $O(a^2)$ calculation in Section~\ref{sec:21} and turn then to the calculation
based on massive OMEs in Section~\ref{sec:22}. Finally, we derive numerical results on the $O(a^2)$ corrections
for a series of key processes to be measured at future facilities.
%--------------------------------------------------------------------------------------------------------
\subsection{Two--loop Corrections: the Direct Calculation}
\label{sec:21}
%--------------------------------------------------------------------------------------------------------

\vspace*{1mm}
\noindent
Upon neglecting power corrections of $O(m_e^2/s)$ the initial state QED 
corrections can be written in terms of the following functions 
%------------------------------------------------------------------------------------------------------------------ 
\begin{eqnarray} 
\label{eq1}
H\left(z,\alpha,\frac{s}{m^2}\right) &=& \delta(1-z) + \sum_{k=1}^\infty \left(\frac{\alpha}{4\pi}\right)^k 
C_k\left(z, \frac{s}{m^2}\right) \\ 
\label{eq2}
C_k\left(z, \frac{s}{m^2}\right) &=& \sum_{l=0}^k \ln^{k-l}\left(\frac{s}{m^2}\right) 
c_{k,l}(z),
\end{eqnarray} 
%------------------------------------------------------------------------------------------------------------------ 
which yield 
the respective differential cross sections by 
%------------------------------------------------------------------------------------------------------------------ 
\begin{eqnarray} 
\frac{d \sigma_{e^+e^-}}{ds'} = \frac{1}{s} \sigma_{e^+e^-}(s') H\left(z,\alpha,\frac{s}{m^2}\right), 
\end{eqnarray} 
%------------------------------------------------------------------------------------------------------------------
with $\sigma_{e^+e^-}(s')$ the scattering cross section without the ISR corrections, 
$\alpha \equiv \alpha(s)$ the fine structure constant and $z = s'/s$, where $s'$ is the invariant mass of the produced 
(off-shell) $\gamma/Z$ boson.

The Born cross section is given by
%-----------------------------------------------------------------------
\begin{eqnarray}
\sigma^{(0)}_{e^+e^-}(s) &=& \frac{4 \pi \alpha^2}{3 s}
                                        N_{C,f} \sqrt{1 - \frac{4 m_f}{s}}
\left[\left(1 + \frac{2 m_f^2}{s} \right) G_1(s)
- 6 \frac{m_f^2}{s} G_2(s)\right], 
%%{\cal G}(s)~,
\end{eqnarray}
%-----------------------------------------------------------------------
see e.g. \cite{BDJ}, where $N_{C,f}$ denotes the color factor for the final state fermions, i.e. $N_C = 3$ for quarks
and $N_C = 1$ for leptons. The 
effective couplings $G_i(s)|_{i=1...2}$ read
%-----------------------------------------------------------------------
\begin{eqnarray}
G_1(s) &=& Q_e^2 Q_f^2 + 2 Q_e Q_f v_e v_f {\sf Re}[\chi_Z(s)]
          +(v_e^2+a_e^2)(v_f^2+a_f^2)|\chi_Z(s)|^2,\\
G_2(s) &=& (v_e^2+a_e^2) a_f^2 |\chi_Z(s)|^2, 
%\\
%%G_3(s) &=& 2 Q_e Q_f a_e a_f {\sf Re}[\chi_Z(s)] + 4 v_e v_f a_e a_f |\chi_Z(s)|^2,
\end{eqnarray}
%-----------------------------------------------------------------------
where the  reduced $Z$--propagator is given by
%-----------------------------------------------------------------------
\begin{eqnarray}
\chi_Z(s) = \frac{s}{s-M_Z^2 + i M_Z \Gamma_Z},
\end{eqnarray}
%-----------------------------------------------------------------------
with $M_Z$ and  $\Gamma_Z$ the mass and the with of the $Z$--boson and 
$m_f$ is the mass of the final state fermion. 
$Q_{e,f}$ are the electromagnetic charges of the electron $(Q_e = -1)$
and the final state fermion, respectively, and
the electro--weak couplings $v_i$ and $a_i$ are given by
%-----------------------------------------------------------------------
\begin{eqnarray}
v_e &=& \frac{1}{\sin\theta_w \cos\theta_w}\left[I^3_{w,e} - 2 Q_e
\sin^2\theta_w\right],\\
a_e &=& \frac{1}{\sin\theta_w  \cos\theta_w} I^3_{w,e}, \\
v_f &=& \frac{1}{\sin\theta_w \cos\theta_w}\left[I^3_{w,f} - 2 Q_f
\sin^2\theta_w\right],\\
a_f &=& \frac{1}{\sin\theta_w  \cos\theta_w} I^3_{w,f}~,
\end{eqnarray}
%-----------------------------------------------------------------------
where $\theta_w$ is the weak mixing angle, and $I^3_{w,i} = \pm 1/2$ the third component
of the weak isospin for up and down particles, respectively.

One may distinguish four processes: {\it i)} photon radiation, {\it ii)} $e^+e^-$ pair emission, {\it iii)} fermionic
pure singlet corrections, and {\it iv)} the interference terms between the diagrams contributing to {\it ii)} and {\it 
iii)}.
Furthermore, there is also $\mu^+\mu^-$ initial state radiation and that of light quarks. The latter contribution
belongs to the QCD corrections, and we will label the one of $\mu^+\mu^-$ pair emission as process {\it v)}. It has 
been described in Refs.~\cite{Berends:1987ab,Kniehl:1988id} correctly, as has been confirmed in 
Ref.~\cite{Blumlein:2019srk,Blumlein:2020jrf}. We will consider the processes {\it i)--iv)} in the following.

The phase--space integrals to be performed are organized as iterated integrals over a set of letters,
which are found by solving associated differential equations, cf. \cite{RAAB1}. The alphabet consists of
letters forming generalized harmonic polylogarithms \cite{Ablinger:2013cf}, but also of elliptic letters.
Since all integrations are indefinite, still iterative integrals are obtained. We therefore call the 
corresponding integrals Kummer--elliptic integrals. The respective iterative integrals have the structure 
%------------------------------------------------------------------------------------------------------------------ 
\begin{eqnarray} 
H_{b,\vec{a}}(z) = \int_0^z dy f_b(y) H_{\vec{a}}(y).
\end{eqnarray} 
%------------------------------------------------------------------------------------------------------------------ 
Examples for these letters are, 
cf.~\cite{Blumlein:2020jrf}
{(using the abbreviation $f_b(t;z,\rho) \equiv f_b$)},
%------------------------------------------------------------------------------------------------------------------ 
\begin{eqnarray} 
      d_{24} &=& \frac{1}{\big(t^2(1-z)^2 - 8 \rho(1+z)t + 4 t z + 16 \rho^2 \big)
\sqrt{t^2 (1-z)^2-8 \rho  t (1+z)+16 \rho ^2}},
\\
      d_{26} &=& \frac{1}{\sqrt{1-t} \big(t^2(1-z)^2 - 8 \rho(1+z)t + 4 t z + 16 \rho^2 \big)
\sqrt{t^2 (1-z)^2-8 \rho  t (1+z)+16 \rho ^2}}.
\end{eqnarray} 
%------------------------------------------------------------------------------------------------------------------ 
These integrals depend on the tiny parameter $\rho = m_e^2/s$ and one may now expand in $\rho$ controlling the accuracy
by high precision numerics for the unexpanded integrals at each step. One finally arrives at iterated 
integrals which integrate in terms
of harmonic polylogarithms or even classical polylogarithms and Nielsen integrals 
\cite{NIELSEN,KMR70,Kolbig:1983qt,Devoto:1983tc,LEWIN1,LEWIN2}.

The size of the amplitudes to be dealt with amounts to 10~Gb (process {\it i)}), 25~kb (process {\it ii)}), 56~kb 
(process {\it iii)}) 
and 124~kb (process {\it iv)}), requiring several months for code design and  30h of computation time. The 
calculation
could only be done by using computer--algorithmic methods, which have been developed very recently.

The 2--loop corrections for process {\it i)} consist out of soft (S), virtual (V) and hard (H) corrections and their
interference terms \cite{Blumlein:2020jrf} which can be written schematically by
%------------------------------------------------------------------------------------------------------------------ 
\begin{eqnarray} 
R^{\gamma\gamma}_2 = 
  T_2^{\rm S_2} 
+ T_2^{\rm V_2} 
+ T_2^{\rm S_1 V_1} 
+ T_2^{\rm S_1 H_1} 
+ T_2^{\rm V_1 H_1} 
+ T_2^{\rm H_2} 
. 
\end{eqnarray} 
%------------------------------------------------------------------------------------------------------------------
We agree with \cite{Berends:1987ab} for all but the virtual--hard contributions, i.e. $T_2^{\rm V_1 H_1}$.
The radiator $R_2^{\gamma\gamma}$ is given by
%--------------------------------------------------------------------------------------------------------
\begin{eqnarray}
R^{\it i)}_2(z) &=& R_2^{\gamma\gamma}(z) 
\nonumber\\
&=&
                  \delta(1-z) \Bigl\{
                        32 (L-1)^2 \ln^2(\varepsilon)
                        + \big( 
                              48 L^2 - (112-64\zeta_2) L + 64 - 64 \zeta_2
                        \big) \ln(\varepsilon)
\nonumber \\ &&
                        + (18-32\zeta_2) L^2
                        - (45-88\zeta_2-48\zeta_3) L
                        + 76 + (6-96\ln(2))\zeta_2 -72 \zeta_3 - \frac{96}{5} \zeta_2^2
                  \Bigr\}
\nonumber \\ &&
                  + \theta(1-z-\varepsilon) \Bigl\{
                        64 (L-1)^2 {\cal D}_1
                        + \big(
                              48 L^2 - (112-64\zeta_2) L + 64 - 64 \zeta_2
                        \big) {\cal D}_0
\nonumber \\ &&
                        - L^2 \biggl(
                               8 (5+z)
                              +32 (1+z) \ln (1-z)
                              +\frac{8 \big(1+3 z^2\big)}{1-z} \ln(z)
                        \biggr)
                        + L \biggl(
                              8 (14+z)
\nonumber \\ &&
                              +\frac{8 \big(5+2 z+7 z^2\big)}{1-z}  \ln(z)
                              -\frac{4 \big(1+3 z^2\big)}{1-z}  \ln^2(z)
                              +\frac{16 \big(1+z^2\big)}{1-z} \text{Li}_2(1-z)
\nonumber \\ &&
                              -32 (1+z) \zeta_2
                              +\biggl[
                                    64 (1+z)
                                    +\frac{16 \big(1+z^2\big)}{1-z}  \ln(z)
                              \biggr] \ln(1-z)
                        \biggr)
                        -\frac{8 \big(18+z-15 z^2\big)}{3 (1-z)}
\nonumber \\ &&
                        -\frac{8 \big(1+z^2\big)}{3 (1-z)}  \ln^3(z)
                        +\frac{4}{3 (1-z)^3} \big(12-33 z+51 z^2-51 z^3+13 z^4\big) \ln^2(z)
\nonumber \\ &&
                        +\frac{32}{3} (3+8 z) \zeta_2
                        -32 (1+z) \zeta_3
                        -\biggl(
                               16 (1+3 z)
                              +\frac{8 \big(2+6 z-3 z^2\big) }{1-z} \ln(z)
\nonumber \\ &&
                              -\frac{8 \big(3-z^2\big)}{1-z}  \ln^2(z)
                        \biggr) \ln(1-z)
                        +16 z \ln^2(1-z)
                        -\biggl(
                               \frac{8 \big(6+3 z+26 z^2-27 z^3\big)}{3 (1-z)^2}
\nonumber \\ &&
                              +\frac{16 \big(1-3 z^2\big)}{1-z} \zeta_2
                        \biggr) \ln(z)
                        +\biggl(
                              -\frac{8 \big(9+19 z-13 z^2\big)}{3 (1-z)}
                              -16 (1+z) \ln(1-z)
\nonumber \\ &&
                              +\frac{8 \big(5-3 z^2\big)}{1-z} \ln(z)
                        \biggr) \text{Li}_2(1-z)
                        +24 (1+z) \text{Li}_3(1-z)
                        +32 (1+z) \text{Li}_3(z)
                  \Bigr\},
\end{eqnarray}
%--------------------------------------------------------------------------------------------------------
with 
%--------------------------------------------------------------------------------------------------------
\begin{eqnarray}
{\cal D}_k(z) = \left(\frac{\ln^{k}(1-z)}{1-z}\right)_+.
\end{eqnarray}
%--------------------------------------------------------------------------------------------------------
Here the separation parameter $\ep$ has been introduced to realize the $+$-distributions in terms 
of functions. It disappears after the $z$-integral has been performed.

Another radiator is the one of the non--singlet process for $e^+e^-$ emission, $R^{\it ii)}_2$.
The correction to the scattering cross section for electron pair emission is given by
%---------------------------------------------------------------------------------------------------
\begin{align}
        R_2^{\it ii)}(z) &=
%%%        \frac{\sigma^{(0)}(s^\prime)}{s} \, a^2 \,
        \Biggl\{
         \frac{64}{3} z (1-z) (1+z-4 \rho) \HT_{v_4,d_7} %\text{Hw3d7}
        +\frac{256}{3} z \rho  (1 + z - 4 \rho) \HT_{v_4,d_6} %\text{Hw3d6}
\nonumber \\ &
        +\frac{128 z (1 - 4 \rho^2 ) (1-z+2 \rho) (1-z-4 \rho)}{3 (1-z)^2} \HT_{d_8,d_7} %\text{Hd8d7}
\nonumber \\ &
        +\frac{512 z \rho  ( 1 - 4 \rho^2 ) (1-z+2 \rho) (1-z-4 \rho)}{3(1-z)^3} \HT_{d_8,d_6} %\text{Hd8d6}
\nonumber \\ &
        +\frac{16}{9 (1-z)^2}
        \Bigl[
                (1+z)^2 \big(4-9 z+4 z^2\big) +2 \big(9-16 z+13 z^2-2 z^3\big) \rho +32 \rho ^2
        \Bigr] \HT_{d_2} %\text{Hd2}
\nonumber \\ &
        +\frac{512 z \rho}{9 (1-z)^4}
        \Bigl[
                3 (1-z)^4 z
                - (1-z)^3 \big(4+z^2\big) \rho
                -2 \big(9-29 z+38 z^2-17 z^3+3 z^4\big) \rho^2
\nonumber \\ &
                -4 (2-z) \big(3+6 z-5 z^2\big) \rho^3
                +16 \big(7-8 z+9 z^2\big) \rho^4
                +128 (3-z) \rho^5
        \Bigr] \HT_{d_4} %\text{Hd4}
\nonumber \\ &
        -\frac{16}{9 (1-z)^4}
        \Bigl[
                3-34 z+129 z^2-212 z^3+129 z^4-34 z^5+3 z^6
                + 8 \big(2-16 z+9 z^2
\nonumber \\ &
                +4 z^3-5 z^4+2 z^5\big) \rho
                +16 z \big(12-13 z+18 z^2-z^3\big) \rho ^2
                +32 \big(1+22 z-7 z^2\big) \rho ^3
        \Bigr] \HT_{d_1} %\text{Hd1}
\nonumber \\ &
        -\frac{128 z}{9 (1-z)^4}
        \Bigl[
                1+7 z-47 z^2+86 z^3-47 z^4+7 z^5+z^6
                - 2 \big(7-55 z+54 z^2
\nonumber \\ &
                +16 z^3-17 z^4+3 z^5\big) \rho
                - 4 \big(39-16 z+16 z^2+4 z^3+5 z^4\big) \rho ^2
\nonumber \\ &
                +16 \big(8-23 z+22 z^2+9 z^3\big) \rho ^3
                +128 \big(7+2 z-z^2\big) \rho ^4
        \Bigr] \HT_{d_5} %\text{Hd5}
        -\frac{64}{3} (2 z+(1-z) \rho) \HT_{d_3} %\text{Hd3}
\nonumber \\ &
        + \biggl[ \frac{16}{3 \sqrt{1-4 \rho }} (1+z-4 \rho) \HT_{v_4} %\text{Hw3}
                + \frac{32 (1-4 \rho^2 ) (1-z+2 \rho) (1-z-4 \rho)}{3 (1-z)^3 \sqrt{1-4 \rho }} \HT_{d_8} %\text{Hd8}
        \biggr]
\nonumber \\ & \times
        \ln \left( \frac{1-z-4 \rho-\sqrt{1-4 \rho } \sqrt{(1-z)^2-8 (1+z) \rho+16 \rho ^2}}{1-z-4 \rho+\sqrt{1-4 \rho }
\sqrt{(1-z)^2-8 (1+z) \rho+16 \rho ^2}} \right)
        \Biggr\}.
\label{eq:DY-NS-FULL}
\end{align}
%---------------------------------------------------------------------------------------------------
It can even be obtained without expanding in the parameter $\rho$. Here the iterative integrals $\HT$ read
%---------------------------------------------------------------------------------------------------
\begin{eqnarray}
\HT_{b,\vec{a}}(z) = \int_z^1 dy f_b(y) \HT_{\vec{a}}(y)~
\end{eqnarray}
%---------------------------------------------------------------------------------------------------
and the respective letters are given in Ref.~\cite{Blumlein:2020jrf}. 
%%%%%%%%%%%%%%%%%%%%%%%%%%%%%%%%%%%%%%%%%%%%%%%%%%%%%%%%%%%%%%%%%%%%%%%%
\begin{figure}[H]
  \centering
  \hskip-0.8cm
  \includegraphics[width=.6\linewidth]{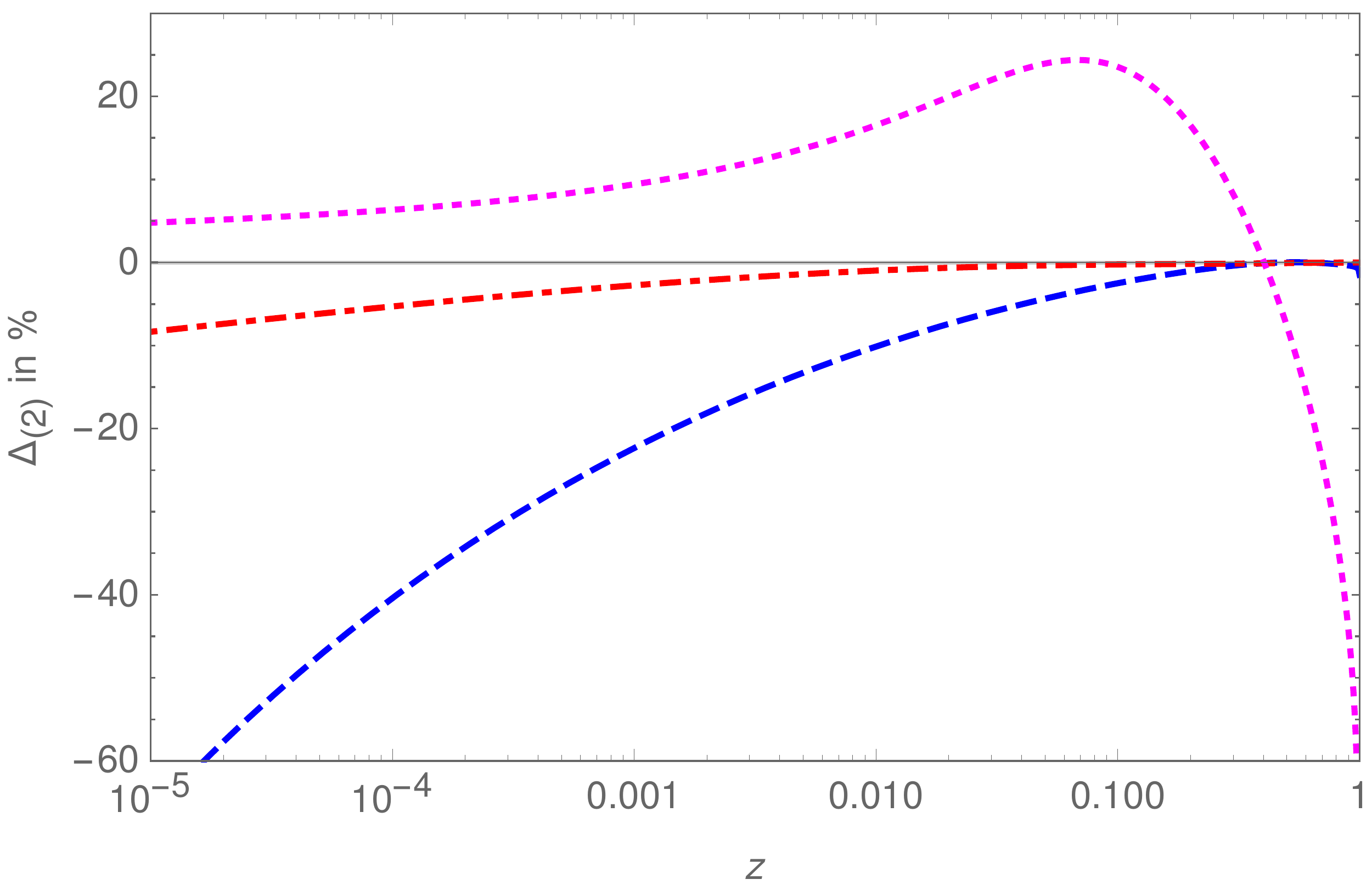}
  \caption{\sf Relative deviations of the results of Ref.~\cite{Berends:1987ab} from the exact result in \% for the 
$O(\alpha^2)$ 
  corrections. 
  The non--singlet contribution (process II): dash-dotted line;
  the pure singlet contribution (process III): dashed; 
  the interference term between both contributions (process IV): dots; for $s = M_Z^2$, 
$M_Z = 91.1879$~GeV; from \cite{Blumlein:2019srk}.} 
  \label{fig:reldev}
\end{figure}
%%%%%%%%%%%%%%%%%%%%%%%%%%%%%%%%%%%%%%%%%%%%%%%%%%%%%%%%%%%%%%%%%%%%%%%%

\noindent
After the expansion in $\rho$ one obtains
%----------------------------------------------------------------------------------------------------------------------------
\begin{align}
\label{eq:R2ee1}
%%R_2^{e^+e^-,\rm NS, \rm H}
%%        \frac{d \sigma^{(2),\text{II}} (z,\rho) }{d \sp }
%%        \frac{\sigma^{(0)}(s^\prime)}{s} \, \left( \frac{\alpha}{4\pi} \right)^2 \,
R_2^{\it ii)}(z) 
&=
 \frac{8 \big(1+z^2\big)}{3 (1-z)} L^2
 +\biggl[
        -\frac{16 \big(11-12 z+11 z^2\big)}{9 (1-z)}
        +\frac{32 \big(1+z^2\big)}{3 (1-z)} \ln(1-z)
\nonumber \\ &
        -\frac{16 \big(1+z^2\big)}{3 (1-z)}  \ln(z)
\biggr] L
+\frac{32 \big(1+z^2\big)}{3 (1-z)}  \ln^2(1-z)
-\frac{16 z^2}{3 (1-z)}  \text{Li}_2(1-z)
\nonumber \\ &
+\frac{32}{9 (1-z)^3} \big(7-13 z+8 z^2-13 z^3+7 z^4\big)
-\frac{16 z}{9 (1-z)^4} \big(3-36 z+94 z^2
\nonumber \\ &
-72 z^3+19 z^4\big)  \ln(z)
-\frac{32 \big(1+z^2\big)}{3 (1-z)} \zeta_2
-\biggl(
        \frac{32 \big(11-12 z+11 z^2\big)}{9 (1-z)}
\nonumber \\ &
        +\frac{32 \big(1+z^2\big)}{3 (1-z)}  \ln(z)
\biggr) \ln(1-z)
-\frac{8 z^2}{3 (1-z)}  \ln^2(z)
        + \mathcal{O}\left( \rho L^2 \right).
\end{align}
%---------------------------------------------------------------------------------------------------
This expression differs from those given in \cite{Berends:1987ab,Kniehl:1988id}.
The other radiators are given in Ref.~\cite{Blumlein:2020jrf}.

Let us compare the fermionic corrections to the non--logarithmic contributions at $O(a^2)$ for processes {\it ii)--iv)} of 
Refs.~\cite{Berends:1987ab} and 
\cite{Blumlein:2019srk,Blumlein:2020jrf}. In Figure~\ref{fig:reldev} we show the relative deviation of the former result 
from the latter.

Both the non--singlet and pure singlet terms grow towards small values of $z$. In the latter case this is caused
also by a deviation in the $1/z$ contributions in \cite{Berends:1987ab}. The difference in the interference term, showing 
various structural differences, first grows and then drops in the large $z$ region, taking negative values.

\begin{figure}[H]
  \centering
%  \hskip-0.8cm
  \includegraphics[width=.6\linewidth]{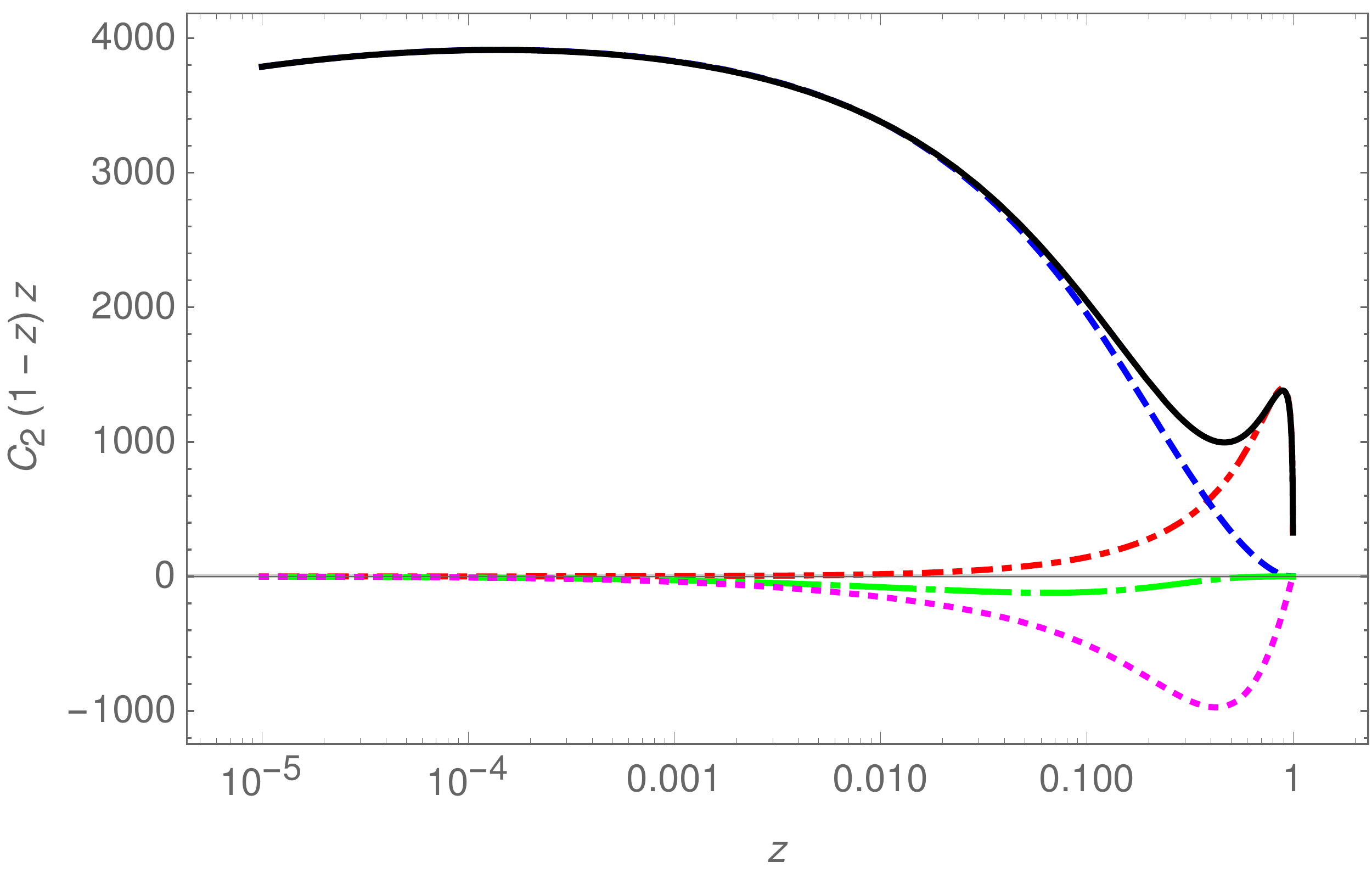}
  \caption{\sf The initial state $O(\alpha^2)$ corrections to $\gamma^*/Z^*$ production due to $e^+ e^-$ pair production 
multiplied by 
$z(1-z)$. 
  The non--singlet contribution (process II): dash-dotted line;
  the pure singlet contribution (process III): dashes; 
  the interference term between both contributions 
  (process IV) $\times 10$: dotted; the vector contributions implied by process B, Ref.~\cite{Hamberg:1990np}, and interferences 
$\times 100$: 
long 
  dash-dotted; all contributions: full line for $s = M_Z^2$; from \cite{Blumlein:2019srk}.} 
  \label{fig:fermcorr}
\end{figure}

Figure~\ref{fig:fermcorr} shows the non--logarithmic contributions to the radiators at $O(a^2)$, weighted by the function
$z(1-z)$ for convenience, to illustrate the relative impact of the contributions from processes {\it ii)--iv)}.
%%%%%%%%%%%%%%%%%%%%%%%%%%%%%%%%%%%%%%%%%%%%%%%%%%%%%%%%%%%%%%%%%%%%%%%%
%%%%%%%%%%%%%%%%%%%%%%%%%%%%%%%%%%%%%%%%%%%%%%%%%%%%%%%%%%%%%%%%%%%%%%%%

\noindent
In the small $z$ region up to about $z \sim 0.2$ the pure singlet contribution dominates. For still larger values,
the non--singlet contribution dominates and at intermediate values there is a negative contribution due to the interference
terms of both. There is another contribution at  $O(a^2)$ having no logarithmic parts, which has not been 
considered
in \cite{Berends:1987ab} but in \cite{Hamberg:1990np}, which is rather small. We would also like to mention that the 
radiative corrections are different for the 
vector--axial vector interference contributions 
if compared to the pure vector and 
pure axial vector contributions, 
which has been considered in Refs.~\cite{Blumlein:2019srk,Blumlein:2020jrf}.

In this and the subsequent calculations reported in this survey we have extensively used the 
packages {\tt FORM} \cite{FORM}
{\tt Sigma} \cite{SIG1,SIG2}, {\tt EvaluateMultiSum} \cite{EMSSP}, {\tt HarmonicSums} \cite{
HARMSU, 
Blumlein:2003gb,
Ablinger:2011te, 
Ablinger:2013cf, 
Ablinger:2014bra,
Vermaseren:1998uu,
Blumlein:1998if,
Remiddi:1999ew,
Blumlein:2009ta}, {\tt HolonomicFunctions} \cite{KOUTSCHAN},
and private implementations \cite{RAAB1} to calculate the respective integrals analytically.
%--------------------------------------------------------------------------------------------------------
\subsection{The Method of Massive Operator Matrix Elements}
\label{sec:22}
%--------------------------------------------------------------------------------------------------------

\vspace*{1mm}
\noindent
The method of massive operator matrix elements has been introduced in QCD to calculate heavy flavor 
corrections to deep--inelastic scattering at two--loop order for all but contributions due to power 
corrections of $O((m^2_Q/Q^2)^k), k \geq 1$, with $Q^2$ the virtuality of the exchanged gauge boson and
$m_Q$ the heavy quark mass \cite{Buza:1995ie}. It is also applicable to massive Drell--Yan processes
as studied in the present paper and has been applied in \cite{Berends:1987ab}, showing that it works 
at one--loop order and for all logarithmic contributions at two--loop order\footnote{Upon correcting 
typographical errors in \cite{Berends:1987ab} in Ref.~\cite{Blumlein:2011mi}.}. However, there has not been a clear
conclusion in \cite{Berends:1987ab}, whether it works also for the non--logarithmic contribution at 
$O(a^2)$. It has been found in \cite{Blumlein:2011mi}, that the $O(a^2)$ terms given in 
\cite{Berends:1987ab} 
are indeed not reproduced. For us it initially caused doubts on its inapplicability, because this 
factorization might not be present in the massive Drell--Yan process, although it is expected.
The complete result obtained in Ref.~\cite{Blumlein:2019qze}, see Section~\ref{sec:21}, has then 
fully confirmed our earlier result in the factorized approach \cite{Blumlein:2011mi}. In the following 
we will outline the different steps in this approach in detail, also for later use at even higher
orders \cite{Ablinger:2020qvo}, see Section~\ref{sec:3}.

The general decomposition of the scattering cross section in Mellin space is given by, 
cf.~\cite{Blumlein:2011mi}\footnote{In the massless case the principle solution of the renormalization group 
equations (RGEs) to general 
orders has been known for long, see~\cite{Blumlein:1997em,Ellis:1993rb}.}
%--------------------------------------------------------------------------------------------------------
\begin{eqnarray}
\label{EQ:SI2}
\frac{d\sigma_{e^+e^-}}{ds'} &=& \frac{1}{s} \sigma^{(0)}(s') 
\Biggl[ \Gamma_{e^+e^+}\left(N,\frac{\mu^2}{m_e^2}\right)
\tilde{\sigma}_{e^+e^-}\left(N,\frac{s'}{\mu^2}\right)
\Gamma_{e^-e^-}\left(N,\frac{\mu^2}{m_e^2}\right)
\nonumber\\
&& +
\Gamma_{\gamma e^+}\left(N,\frac{\mu^2}{m_e^2}\right)
\tilde{\sigma}_{e^- \gamma}\left(N,\frac{s'}{\mu^2}\right)
\Gamma_{e^-e^-}\left(N,\frac{\mu^2}{m_e^2}\right)
\nonumber\\
&& +
\Gamma_{e^+ e^+}\left(N,\frac{\mu^2}{m_e^2}\right)
\tilde{\sigma}_{e^+ \gamma}\left(N,\frac{s'}{\mu^2}\right)
\Gamma_{\gamma e^-}\left(N,\frac{\mu^2}{m_e^2}\right)
\nonumber\\
&& +
\Gamma_{\gamma e^+}\left(N,\frac{\mu^2}{m_e^2}\right)
\tilde{\sigma}_{\gamma \gamma}\left(N,\frac{s'}{\mu^2}\right)
\Gamma_{\gamma e^-}\left(N,\frac{\mu^2}{m_e^2}\right)
\Biggr].
\end{eqnarray}
%--------------------------------------------------------------------------------------------------------
The terms in the brackets $[...]$ are Mellin--convoluted. Only massive OMEs of the kind 
$\Gamma_{e^{\pm} e^{\pm}}$ and $\Gamma_{\gamma e^{\pm}}$ 
contribute because the process considered has electron--positron initial states and the 
$\tilde{\sigma}_{\gamma\gamma}$ terms appear only from $O(a^4)$ onward. 
{In the factorization also the massless Wilson coefficients for the 
Drell-Yan process $\tilde{\sigma}_{ij}$, $i,j=e^{\pm},\gamma$ contribute.} 
The factorization scale $\mu$ cancels after expanding in $a_0 =  a(m_e^2)$.
The following expansions hold for the massive OMEs and massless Wilson coefficients
%--------------------------------------------------------------------------------------------------------
\begin{eqnarray}
\Gamma_{li}\left(N, \frac{\mu^2}{m_e^2}\right)       &=& \delta_{li} + \sum_{r=1}^\infty a^r(\mu^2) \sum_{n=0}^r 
a_{li;nr}(N) \Lambda^n
\\
\tilde{\sigma}_{lk}\left(N, \frac{s'}{\mu^2} \right) &=& \delta_{lk} + \sum_{r=1}^\infty a^r(\mu^2) \sum_{n=0}^r 
b_{lk;nr}(N) 
\lambda^n,
\end{eqnarray}
%--------------------------------------------------------------------------------------------------------
with the logarithms $\Lambda$ and $\lambda$ given by
%--------------------------------------------------------------------------------------------------------
\begin{eqnarray}
\Lambda        = \ln\left(\frac{\mu^2}{m_e^2}\right),~~~~~~~
\lambda  = \ln\left(\frac{s'}{\mu^2}\right).
\end{eqnarray}
%--------------------------------------------------------------------------------------------------------
The massive OMEs $\Gamma_{ij}$ and massless Wilson coefficients $\tilde{\sigma}_{kl}$ fulfill the 
following renormalization group equations, cf.~\cite{Blumlein:2000wh}, 
%--------------------------------------------------------------------------------------------------------
\begin{eqnarray}
\label{eq:RE1}
\left[\left(\frac{\partial}{\partial \Lambda} + \beta(a) \frac{\partial}{\partial a}\right) \delta_{kl} + 
\frac{1}{2} 
\gamma_{kl}(N,a)\right]
\Gamma_{li}\left(N,a,\frac{\mu^2}{m_e^2}\right) 
&=& 0 \\
%----
\label{eq:RE2}
\left[\left(\frac{\partial}{\partial \lambda} - \beta(a) \frac{\partial}{\partial a}\right) \delta_{kl} \delta_{jm} 
+ \frac{1}{2} \gamma_{kl}(N,a) \delta_{jm}
+ \frac{1}{2} \gamma_{jm}(N,a) \delta_{kl} \right]
\tilde{\sigma}_{lj}\left(N,a,\frac{s'}{\mu^2}\right) &=& 0~, 
\end{eqnarray}
%--------------------------------------------------------------------------------------------------------
and the QED $\beta$ function has the representation 
%--------------------------------------------------------------------------------------------------------
\begin{eqnarray}
\beta(a) = - \sum_{k=0}^\infty \beta_k a^{k+2},
\end{eqnarray}
%--------------------------------------------------------------------------------------------------------
with $\beta_0 = -4/3,~\beta_1 = -4$.

The calculation of the scattering cross section (\ref{EQ:SI2}) to $O(a_0^2)$ requires the knowledge 
of the expansion coefficients of the massless Drell--Yan cross section 
\cite{Hamberg:1990np,Harlander:2002wh,Blumlein:2019srk} to $O(a_0^2)$, the one-- and two--loop anomalous dimensions 
$\gamma_{ij}^{(k)}$ \cite{ANOM2}, and the massive OMEs $\Gamma_{ee}^{(0)}, \Gamma_{e\gamma}^{(0)}$ and 
$\Gamma_{ee}^{(1)}$. 
The latter quantity needs to be known in its contributions to processes {\it i) + iv), ii)} and {\it 
iii)}. The major task in Ref.~\cite{Blumlein:2011mi} has been to calculate $\Gamma_{ee}^{(1)}$. At that
time, many of the automated methods and function spaces found in later massive calculations have been 
not yet known, cf.~\cite{Blumlein:2018cms}, and majorly hypergeometric integration techniques \cite{HYP}
were used. Checks have been performed using the package {\tt tarcer} \cite{Mertig:1998vk} and detailed
lists of special integrals had to be calculated,~cf.~\cite{Blumlein:2011mi}.

As an example, we show the massive OME $\hat{\hat{A}}_{ee}^{\rm III}$, from which $\Gamma_{ee}^{\rm III}$
is obtained,
%--------------------------------------------------------------------------------------------------------
\begin{eqnarray}
\hat{\hat{A}}_{ee}^{(2), \rm III}
&=& S_\ep^2 \int^1_0 dx \,\, x^N \frac{1+(-1)^N}{2} \BLCB
\frac{8}{\ep^2} \left[ \frac{1-x}{3 x} (4 x^2+7 x+4)+2 (1+x) \ln(x) \right]
\nonumber \\ && \phantom{ \int^1_0 dx \,\, x^n \BLCB}
+\frac{4}{\ep} \BLB 5 (1+x) \ln^2(x) 
-\frac{1+x}{3 x} (8 x^2-17 x-16) \ln(x)
+\frac{4 (1-x)}{9 x} (5 x^2
\nonumber \\ && \phantom{ \int^1_0 dx \,\, x^n \BLCB}
+23 x+14) \BRB
+\frac{2}{x} (1-x) (4 x^2+13 x+4) \zeta_2
+\frac{1}{3 x} (8 x^3+135 x^2+75 x
\nonumber \\ && \phantom{ \int^1_0 dx \,\, x^n \BLCB}
+32) \ln^2(x)
+\left[ \frac{304}{9 x}-\frac{80}{9} x^2-\frac{32}{3} x+108
-\frac{32}{1+x}-\frac{64 (1+2 x)}{3 (1+x)^3} \right] \ln(x)
\nonumber \\
&& \phantom{ \int^1_0 dx \,\, x^n \BLCB}
-\frac{224}{27} x^2
+16 \frac{1-x}{3 x} (x^2+4 x+1) \left[ 2 \ln(x) \ln(1+x)
-{\rm Li}_2(1-x) \right.
\nonumber \\ && \phantom{ \int^1_0 dx \,\, x^n \BLCB}
\left.
+2 {\rm Li}_2(-x) \right]
+(1+x) \BLB
4 \zeta_2 \ln(x) 
+\frac{14}{3} \ln^3(x)
-32 \ln(x) {\rm Li}_2(-x)
\nonumber \\ && \phantom{ \int^1_0 dx \,\, x^n \BLCB}
-16 \ln(x) {\rm Li}_2(x)
+64 {\rm Li}_3(-x)
+32 {\rm Li}_3(x)
+16 \zeta_3
\BRB 
-\frac{182}{3} x+50
\nonumber \\ && \phantom{ \int^1_0 dx \,\, x^n \BLCB}
-\frac{32}{1+x}
+\frac{800}{27 x}
+\frac{64}{3 (1+x)^2} 
\BRCB m_e^{2\ep} (\Delta \cdot p)^N~.
\label{BigA2IIIunre}
\end{eqnarray}
%--------------------------------------------------------------------------------------------------------
Here $\Delta$ is a light--like vector, $p$ the external momentum of the OME, $N$ labels the Mellin 
moment, $\ep = D-4$ is the dimensional parameter, and $S_\ep = \exp[(\ep/2)(\gamma_E - \ln(4\pi))]$ 
is the spherical factor, with $\gamma_E$ the Euler--Mascheroni constant.

After all building blocks which enter (\ref{EQ:SI2}) have been calculated, one can perform the corresponding
Mellin transforms analytically and obtain all the radiators, which were also obtained in the direct 
calculation, presented in Section~\ref{sec:21}.  
%--------------------------------------------------------------------------------------------------------
\subsection{Numerical Results at \boldmath $O(\alpha^2)$}
\label{sec:23}
%--------------------------------------------------------------------------------------------------------

\vspace*{1mm}
\noindent
The initial state QED corrections can be written in terms of the radiator functions introduced in 
Section~\ref{sec:21}, which we use in the following to calculate the ISR corrections to different
processes. 
%%%%%%%%%%%%%%%%%%%%%%%%%%%%%%%%%%%%%%%%%%%%%%%%%%%%%%%%%%%%%%%%%%%%%%%%
\begin{figure}[H]
  \centering
  \hskip-0.8cm
  \includegraphics[width=.6\linewidth]{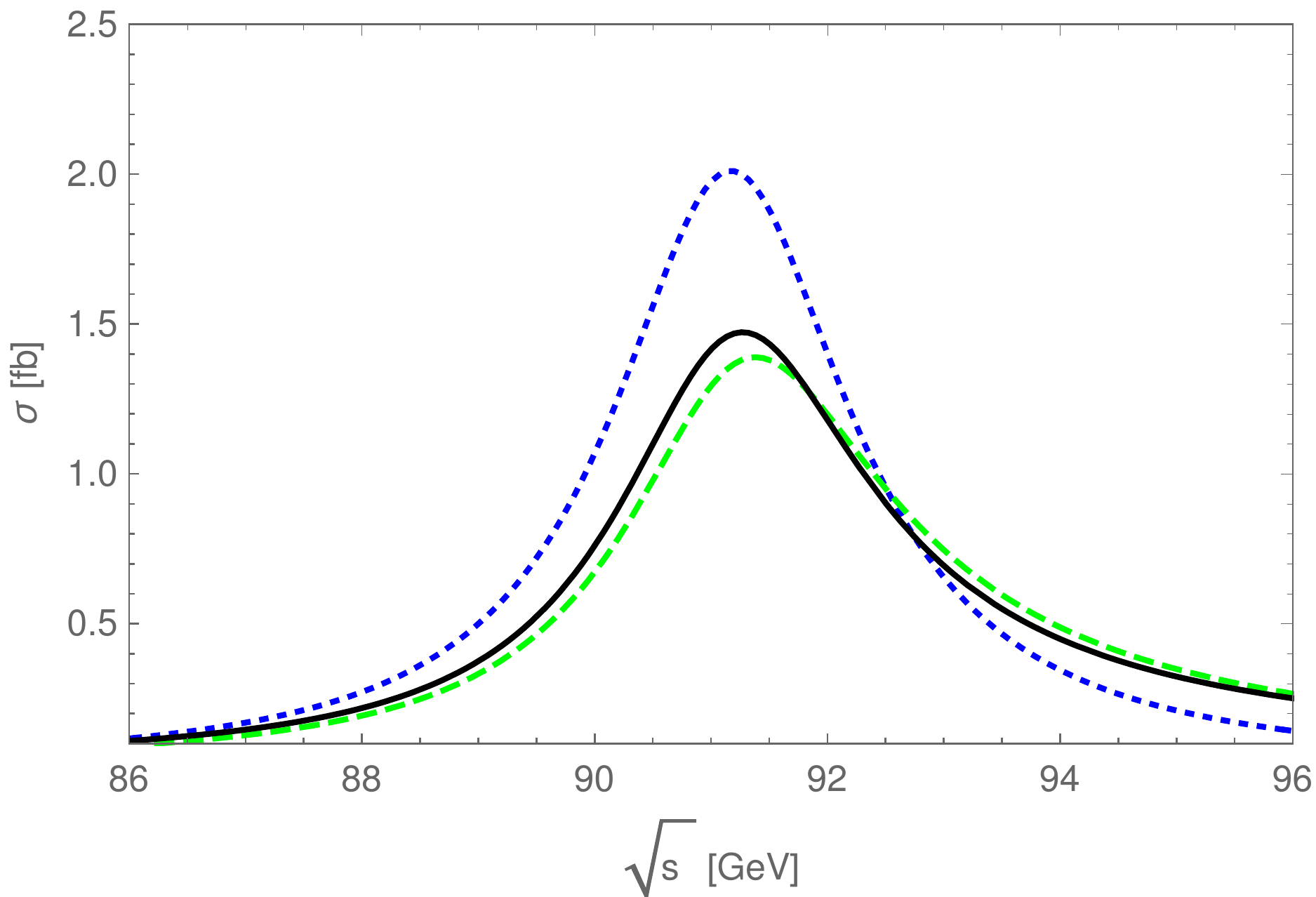}
  \caption[]{\sf The $Z$-resonance in $e^+e^- \rightarrow \mu^+\mu^-$. Dotted line: Born cross section; Dashed line: 
  $O(\alpha)$ ISR corrections; 
  Full line: $O(\alpha^2)$ + soft resummation ISR corrections, with $s_0 = 4 m_\tau^2$; from 
  Ref.~\cite{Blumlein:2019pqb}.} 
  \label{fig:ZPEAK2}
\end{figure}
%%%%%%%%%%%%%%%%%%%%%%%%%%%%%%%%%%%%%%%%%%%%%%%%%%%%%%%%%%%%%%%%%%%%%%%%
%%%%%%%%%%%%%%%%%%%%%%%%%%%%%%%%%%%%%%%%%%%%%%%%%%%%%%%%%%%%%%%%%%%%%%%%
\begin{figure}[H]
  \centering
  \hskip-0.8cm
  \includegraphics[width=.6\linewidth]{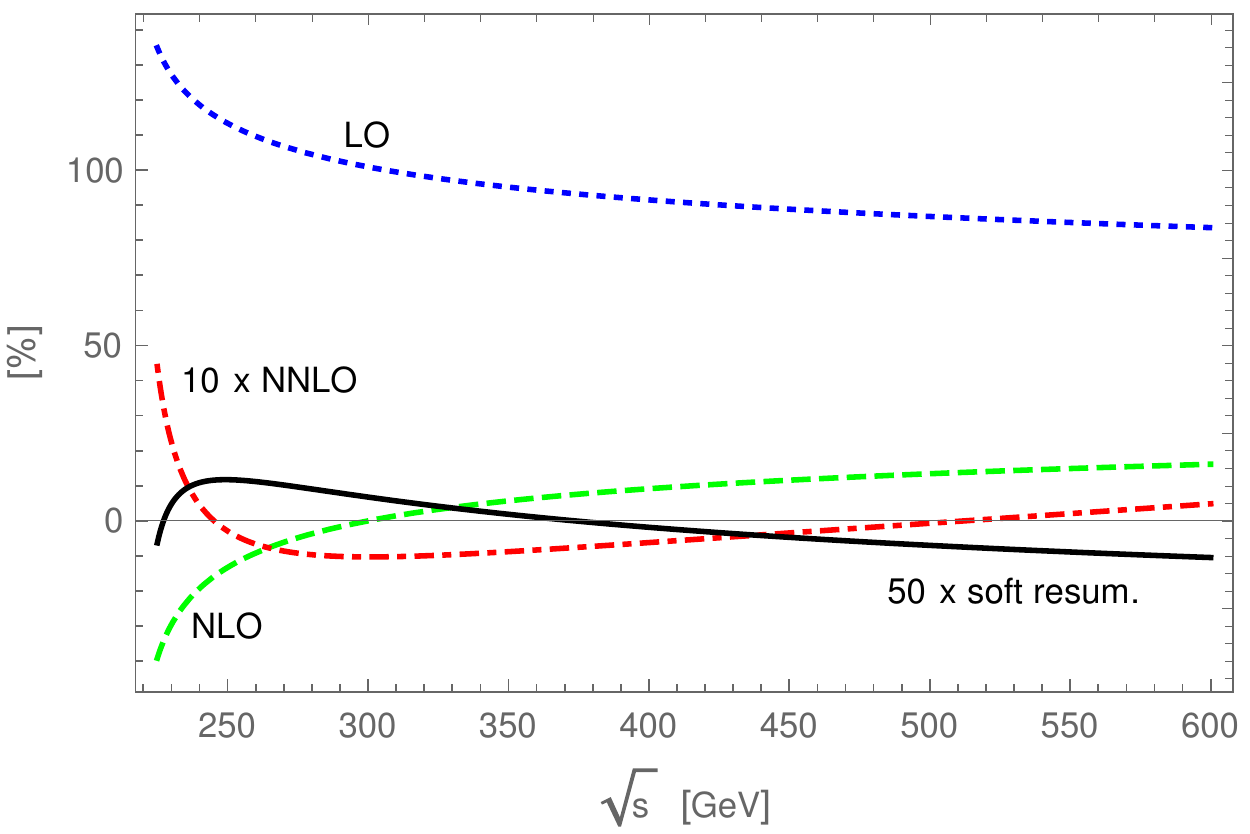}
  \caption[]{\sf Relative contributions of the ISR QED corrections to the cross section for $e^+e^- \rightarrow Z H$ in \%. 
  Dotted line: $O(\alpha^0)$;
  Dashed line: $O(\alpha)$;
  Dash-dotted line: $O(\alpha^2)$;
  Full line: soft resummation beyond $O(\alpha^2)$, with $s_0 = 4 m_\tau^2$; from Ref.~\cite{Blumlein:2019pqb}.}
  \label{fig:ZH3}
\end{figure}
%%%%%%%%%%%%%%%%%%%%%%%%%%%%%%%%%%%%%%%%%%%%%%%%%%%%%%%%%%%%%%%%%%%%%%%%

\noindent
Here we consider the contribution to $O(\alpha^2)$
only and turn to higher order corrections in Section~\ref{sec:3}. We discuss the ISR corrections
for the $Z$ peak and its surrounding, the process $e^+e^- \rightarrow ZH$, and $t\bar{t}$ production
in the threshold region, cf.~Ref.~\cite{Blumlein:2019pqb}.
The ISR corrections depend on the experimental cut $s_0$. In the LEP1 analysis it has been chosen $s_0 = 
4 m_\tau^2$ or $s_0 = 0.01 M_Z^2$ \cite{ALEPH:2005ab}. Furthermore, one considers the cases of a fixed $Z$-width
or the $s$-dependent $Z$-width, leading to a peak shift of $34.2$ MeV and a shift of the width of 1 MeV, in 
accordance with Refs.~\cite{SDEP}. We illustrate the different QED ISR corrections to $e^+e^- 
\rightarrow Z^*/\gamma^*$ around the $Z$ peak in Figure~\ref{fig:ZPEAK2}. The ISR corrections change the profile 
of the resonance, i.e. the peak position, height and the half width. The effect of soft resummation beyond 
$O(\alpha^2)$ are nearly invisible.

More detailed effects, also due to higher order corrections, are discussed in Table~\ref{TAB1} in Section~\ref{sec:3}.  
The new results, compared with  \cite{Berends:1987ab}, imply a relative shift in the $Z$-width by $\sim$4 
MeV for 
$s_0 = 4 m_\tau^2$, being larger than the current value of $\Delta \Gamma_Z = \pm 2.3$ MeV \cite{PDG}, which may 
require 
a reanalysis of the LEP data to obtain consistent results.

For the study of the radiative corrections to the process $e^+e^- \rightarrow Z H$ we refer to the
Born cross section in \cite{Barger:1993wt}. The anticipated experimental accuracy for this process should
reach 1\% \cite{dEnterria:2016sca} at future colliders like the ILC, CLIC, and even $0.4\%$ at the FCC\_ee 
\cite{Ruan:2014xxa}. Figure~\ref{fig:ZH3} shows the relative corrections of ISR corrections.

%%%%%%%%%%%%%%%%%%%%%%%%%%%%%%%%%%%%%%%%%%%%%%%%%%%%%%%%%%%%%%%%%%%%%%%%
\begin{figure}[H]
  \centering
  \hskip-0.8cm
  \includegraphics[width=.6\linewidth]{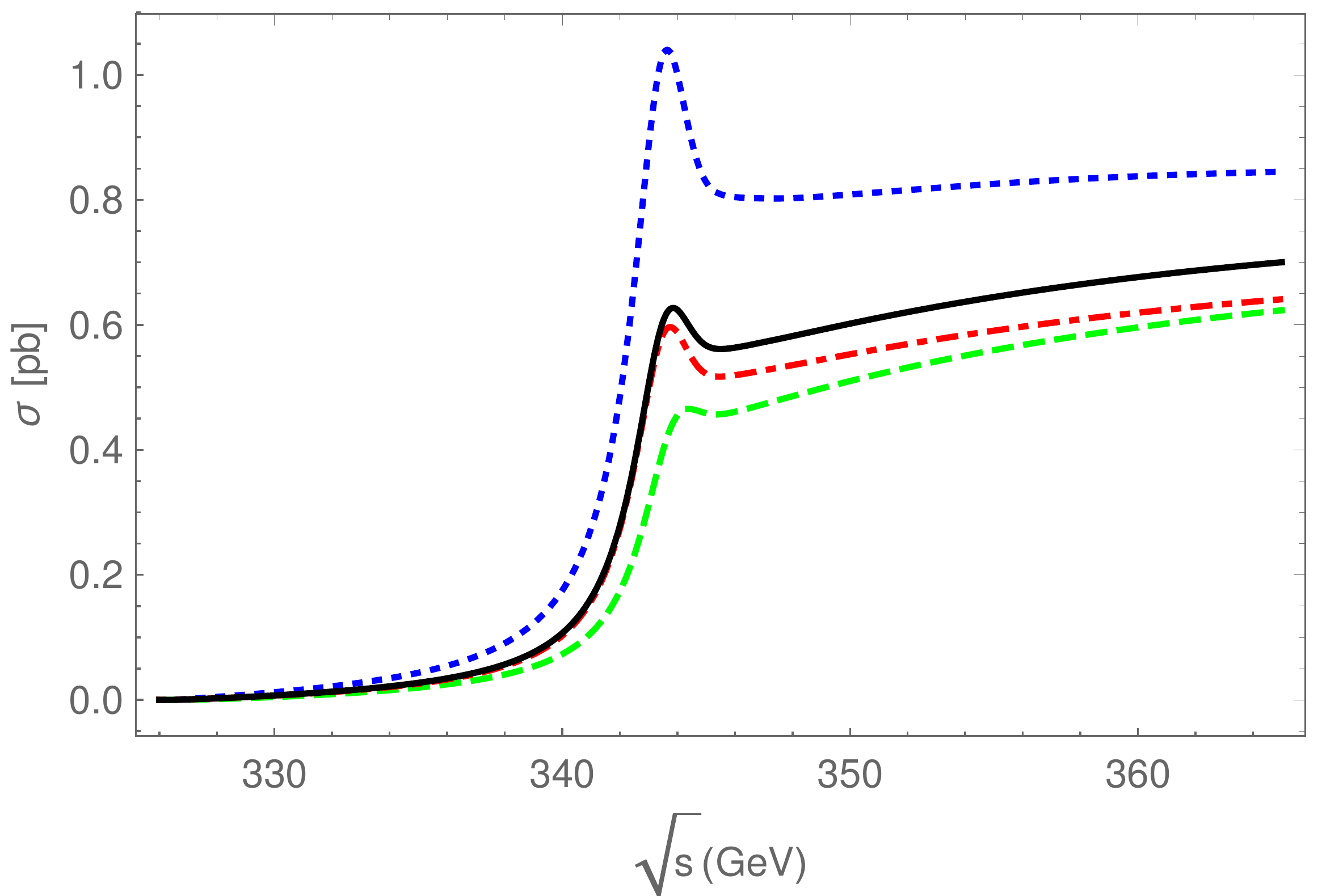}
  \caption[]{\sf The QED  ISR corrections to $e^+e^- \rightarrow t\overline{t}$ ($s$-channel photon exchange) in the threshold 
  region for a PS-mass of $m_t = 172~\GeV$.
Dotted line $O(\alpha^{0})$;
Dashed line $O(\alpha)$;
Dash-dotted line $O(\alpha^{2})$;
Full line $O(\alpha^{2})$ + soft resummation; from Ref.~\cite{Blumlein:2019pqb}.  
\label{fig:TT1}}
\end{figure}
%%%%%%%%%%%%%%%%%%%%%%%%%%%%%%%%%%%%%%%%%%%%%%%%%%%%%%%%%%%%%%%%%%%%%%%%
\noindent
The NNLO corrections vary between +4.8\% and $-1\%$ and are thus larger or of the size of the expected
experimental errors, which might imply the need of higher order corrections. On the other hand,
the soft resummation contributions are of $O(\pm 0.2\%)$ and reach half of the projected accuracy. 

The QED corrections to $t\bar{t}$ production in the threshold region are shown in Figure~\ref{fig:TT1}. For 
the cross section at threshold we use the N$^3$LO QCD corrections implemented in the code {\tt QQbar\_threshold} 
\cite{Beneke:2016kkb,Beneke:2017rdn,Beneke:2015kwa} without QED corrections. Figure~\ref{fig:TT1} illustrates
the convergence of the corrections from the uncorrected cross section to the one with the $O(\alpha^2)$ corrections
and higher order soft resummation, which is stable up to the peak region an displays still some shift towards 
the continuum region. Accuracy studies for this process have been made in Refs.~\cite{Seidel:2013sqa,Simon:2016pwp}.

The full two--loop QED corrections are mandatory down to the constant contribution to match the anticipated
experimental accuracy for key processes to be measured at future $e^+e^-$ colliders and they were mandatory
in the analysis of the LEP experiments. Given the luminosity projected for future experiments, even higher order
corrections will become necessary, to which we turn now. 
%--------------------------------------------------------------------------------------------------------
\section{The Higher Order Corrections}
\label{sec:3}
%--------------------------------------------------------------------------------------------------------

\vspace*{1mm}
\noindent
In the previous section it has been shown, that the method of massive OMEs provides an equivalent way
to calculate the initial state corrections in the limit $m_e^2/s \rightarrow 0$. 
{This is fully justified in particular for inclusive calculations, where the power corrections 
being neglected do not play any role as e.g. in high energy electron-positron collisions. }
However, the constant part is still important. One arrives at hierarchies of the kind
$O(\alpha^k L^k), O(\alpha^k L^{k-1}), O(\alpha^k L^{k-2}), ..., O(\alpha^k)$. By using the 
renormalization group equation one can now derive three consecutive corrections, for which all massive
OMEs are available together with the two--loop corrections to the massless Drell--Yan cross section
\cite{Hamberg:1990np,Harlander:2002wh,Blumlein:2019srk}. In Ref.~\cite{Ablinger:2020qvo} we have 
calculated all the corresponding corrections up to $O(\alpha^6 L^5)$.

{To obtain the leading logarithmic contributions to the radiators}
we solve (\ref{EQ:SI2}) to the desired order and calculate the coefficients of the expansion of the 
inclusive radiator given in Eqs.~(\ref{eq1},\ref{eq2}) in Mellin space.
While at lower orders all terms are obtained, at three--loop order we get the coefficients
$c_{3,3}, c_{3,2}$ and $c_{3,1}$ only, etc. 
{
  For example we obtain
  \begin{eqnarray}
    c_{3,1} &=& 
        -\gamma_{ee}^{(2)}
        -2 \Gamma_{ee}^{(0)} \gamma_{ee}^{(1)}
        - \Gamma_{ee}^{(0)} \gamma_{e \gamma}^{(0)} \Gamma_{\gamma e}^{(0)}
        - \gamma_{e \gamma}^{(1)} \Gamma_{\gamma e}^{(0)}
        - \gamma_{e \gamma}^{(0)} \Gamma_{\gamma e}^{(1)}
        -\beta_1 \sigma_{ee}^{(0)}
        - \gamma_{ee}^{(1)} \sigma_{ee}^{(0)}
        \nonumber\\ & &
        - \gamma_{e \gamma}^{(0)} \Gamma_{\gamma e}^{(0)} \sigma_{ee}^{(0)}
        - 2 \Gamma_{ee}^{(0)} \gamma_{\gamma e}^{(0)} \sigma_{e \gamma}^{(0)}
        - \gamma_{\gamma e}^{(1)} \sigma_{e \gamma}^{(0)}
        - \Gamma_{\gamma e}^{(0)} \gamma_{\gamma \gamma}^{(0)} \sigma_{e \gamma}^{(0)}
        - \gamma_{\gamma e}^{(0)} \sigma_{\gamma e}^{(1)}
        + \beta_0 \Bigl[-2 \Gamma_{ee}^{(0)} \sigma_{ee}^{(0)}
        \nonumber\\ &&
        -2 \sigma_{ee}^{(1)}
        -2 \Gamma_{\gamma e}^{(0)} \sigma_{e \gamma}^{(0)}
        \Bigr]
        - \gamma_{ee}^{(0)} \left[
                 {\Gamma_{ee}^{(0)}}^2
                +2 \Gamma_{ee}^{(1)}
                +2 \Gamma_{ee}^{(0)} \sigma_{ee}^{(0)}
                +\sigma_{ee}^{(1)}
                +\Gamma_{\gamma e}^{(0)} \sigma_{e \gamma}^{(0)} \right].
        \label{eq:c31}
  \end{eqnarray}
}

\noindent
The calculation is best carried out in Mellin space and 
the $z$--space representation is finally obtained by an analytic Mellin inversion using the tools of the 
package {\tt HarmonicSums}. 

{As can be seen from Eq.~(\ref{eq:c31}),}
starting with three--loop order the expansion coefficient $\Gamma_{\gamma e}^{(1)}$
contributes. In Mellin space it is obtained from the massive OME $A_{\gamma e}^{(2)}$, cf.~\cite{Ablinger:2020qvo},
%-----------------------------------------------------------------------------------------------------------------
\begin{eqnarray}
\label{eq:AN}
      A_{\gamma e}^{(2)}(N) &=&
      \Biggl[ 
            \frac{\big(N^2+N+2\big)\big(N^2+N+6\big)}{3 (N-1) N^2 (N+1)^2}
            -\frac{4 \big(N^2+N+2\big)}{(N-1) N (N+1)} S_1
      \Biggr] L^2
\nonumber \\ &&
      - \Biggl[ 
            \frac{2 P_2}{9 (N-1)^2 N^3 (N+1)^3}
            -\frac{4 P_1}{3 (N-1) N^2 (N+1)^2} S_1
            +\frac{12 \big(N^2+N+2\big)}{(N-1) N (N+1)} S_1^2
\nonumber \\ &&
            +\frac{12 \big(N^2+N+2\big)}{(N-1) N (N+1)} S_2
      \Biggr] L
      +\frac{P_ 8}{27 (N-4) (N-3) (N-2) (N-1) N^4 (N+1)^4}
\nonumber \\ &&
      +\biggl(
            \frac{2 P_ 7}{9 (N-4) (N-3) (N-2) (N-1) N^3 (N+1)^3}
            +\frac{2 \big(N^2+N+2\big)}{(N-1) N (N+1)} S_ 2            
      \biggr) S_ 1
\nonumber \\ &&
      +\frac{P_ 3}{3 (N-2) (N-1) N (N+1)^2} S_ 1^2
      +\frac{2 \big(N^2+N+2\big)}{3 (N-1) N (N+1)} S_ 1^3
\nonumber \\ &&
      +\frac{P_ 6}{3 (N-2) (N-1) N^2 (N+1)^2} S_ 2
      +\frac{4 \big(N^2+N+2\big)}{3 (N-1) N (N+1)} S_ 3
\nonumber \\ &&
      -\frac{48 \big(N^2+N+2\big)}{(N-1) N (N+1)} S_{2,1}
      +\frac{3 \cdot 2^{6+N}}{(N-2) (N+1)^2} S_ {1,1}\biggl(\frac{1}{2},1\biggr)
\nonumber \\ &&
      + \frac{2^{6-N} P_ 5}{3 (N-3) (N-2) (N-1)^2 N^2} 
      \biggl(
              S_ 2(2)
            + S_ 1 S_1(2)
            - S_{1,1}(1,2)
            - S_{1,1}(2,1)
      \biggr)
\nonumber \\ &&
      -\frac{32 \big(N^2+N+2\big)}{(N-1) N (N+1)}
      \biggl[
            S_ 1(2) S_ {1,1}\biggl(\frac{1}{2},1\biggr)
            + S_ {1,2}\biggl(\frac{1}{2},2\biggr)
            - S_ {1,1,1}\biggl(\frac{1}{2},1,2\biggr)
\nonumber \\ &&
            - S_ {1,1,1}\biggl(\frac{1}{2},2,1\biggr)
            - \frac{\zeta_2}{2} S_1(2)
      \biggr]
      + \frac{4 P_ 4}{(N-2) (N-1) N^2 (N+1)^2} \zeta_2,
\end{eqnarray}
with the polynomials
\begin{eqnarray}
      P_1 &=& 25 N^4+44 N^3+87 N^2+56 N+12,
\\
      P_2 &=& 112 N^7+194 N^6+347 N^5+339 N^4+93 N^3-293 N^2-60 N+36,
\\ 
      P_3 &=& 17 N^4-66 N^3-179 N^2-272 N-212,
\\
      P_4 &=& N^5+4 N^4 +25 N^3+14 N^2+12 N+8 -3 \cdot 2^{N+3} N^2 \bigl(N - 1\bigr),
\\
      P_5 &=& 9 N^5-24 N^4+8 N^3+4 N^2+33 N-18,
\\
      P_6 &=& 11 N^5-90 N^4-329 N^3-356 N^2-284 N-48,
\\
      P_7 &=& 17 N^9+213 N^8-1729 N^7+2329 N^6-5196 N^5+7898 N^4+16196 N^3
\nonumber \\ &&
      +12528 N^2-4896 N-3456,
\\
      P_8 &=& -509 N^{11}+2365 N^{10}+2797 N^9-13158 N^8+31274 N^7-4694 N^6-64636 N^5
\nonumber \\ &&
      -107861 N^4-14622 N^3+6588 N^2-2376 N-2592.
\end{eqnarray}
%-----------------------------------------------------------------------------------------------------------------

The radiators are obtained in terms of harmonic sums 
\cite{Vermaseren:1998uu,Blumlein:1998if} and generalized harmonic sums \cite{Ablinger:2013cf}
or harmonic polylogarithms \cite{Remiddi:1999ew} $H_{\vec{a}}(z)$ and $H_{\vec{a}}(1-z)$ weighted by 
rational coefficients. The choice of arguments has been made to avoid the occurrence of more general 
iterated integrals.  As usual, the radiators in $z$-space are distribution valued.
We show the coefficient $c_{3,1}$ in Mellin space as an example,
%---------------------------------------------------------------------------------------------------------
\begin{eqnarray}
%---------------------------------------------------------------------------------------------------------
c_{3,1} &=& \Biggl\{
       \frac{\big(
                2+N+N^2\big) 2^{8-N} [S_2(2) - S_{1,1}(1,2) - S_{1,1}(2,1)]  Q_{36}}
{3 (N-3) (N-2) (N-1)^2 N^3 (N+1) (N+2)}
\nonumber\\ &&
        +\frac{16 S_{2,1} Q_{54}}{9 (N-1) N^2 (N+1)^2 (N+2)}
        +\frac{8 S_3 Q_{91}}{27 (N-1) N^2 (N+1)^2 (N+2)}
        \nonumber\\
        &&+\frac{Q_{206}
        }{81 (N-4) (N-3) (N-2) (N-1)^2 N^5 (N+1)^5 (N+2)^3}
\nonumber\\ &&     
   +\Biggl[
                \Biggl[
                        -
                        \frac{64 S_1 Q_{119}}{3 (N-1) N^2 (N+1)^2 (N+2)^2}
+\frac{32 Q_{144}}{9 (N-1) N^3 (N+1)^3 (N+2)^2}
                        \nonumber\\
                        &&
                        +\frac{768 (N+3) \big(
                                2+N^2\big) S_1^2}{(N-1) N (N+1) (N+2)}
                \Biggr] S_{-2}
                +\Biggl[
-\frac{32 (N+3) \big(
                                2+N^2
                        \big)
\big(6+13 N+13 N^2\big)}{3 (N-1) N^2 (N+1)^2 (N+2)}
\nonumber\\ &&                        
+\frac{128 (N+3) \big(
                                2+N^2\big) S_1}{(N-1) N (N+1) (N+2)}
                \Biggr] S_{-3}
+\frac{64 (N+3) \big(
                        2+N^2
                \big)
\big(6+13 N+13 N^2\big) S_{-2,1}}{(N-1) N^2 (N+1)^2 (N+2)}
                \nonumber\\
                &&
                -\frac{768 (N+3) \big(
                        2+N^2\big) S_1 S_{-2,1}}{(N-1) N (N+1) (N+2)}
                +\Biggl[
-\frac{32 S_1 Q_{119}}{3 (N-1) N^2 (N+1)^2 (N+2)^2}
\nonumber\\ &&                    
 +\frac{16 Q_{144}}{9 (N-1) N^3 (N+1)^3 (N+2)^2}
    +\frac{384 (N+3) \big(
                                2+N^2\big) S_1^2}{(N-1) N (N+1) (N+2)}
                \Biggr] \zeta_2
\nonumber\\ &&                    
                +\Biggl[
                        \frac{32 (N+3) \big(
                                2+N^2
                        \big)
\big(6+13 N+13 N^2\big)}{(N-1) N^2 (N+1)^2 (N+2)}
                        -\frac{384 (N+3) \big(
                                2+N^2\big) S_1}{(N-1) N (N+1) (N+2)}
                \Biggr] \zeta_3
        \Biggr] (-1)^N
                \nonumber\\
                &&
 +\Biggl[
                -\frac{64 S_{2,1} Q_6}{3 (N-1) N (N+1)}
                +\frac{\big(
                        2+N+N^2\big) 2^{8-N} S_1(2) Q_{36}}{3 (N-3) (N-2) (N-1)^2 N^3 (N+1) (N+2)}
       \nonumber\\ &&         
-
                \frac{64 S_3 Q_{20}}{9 (N-1) N}
-\frac{16 S_2 Q_{132}}{27 (N-1) N^2 (N+1)^2 (N+2)^2}
                \nonumber\\
                &&-\frac{8 Q_{201}}{81 (N-4) (N-3) (N-2) (N-1)^2 N^4 (N+1)^4 (N+2)^3}
                -1152 S_2^2
                -384 S_4
                \nonumber\\
                &&-1280 S_{3,1}
                -1024 S_{-2,-1}
                +768 S_{2,1,1}
        \Biggr] S_1
        +\Biggl[
                \frac{32 S_2 Q_8}{(N-1) N (N+1)}
                -256 S_{2,1}
                \nonumber\\
                &&+\frac{16 Q_{150}}{27 (N-2) (N-1)^2 N^3 (N+1)^3 (N+2)^2}
        \Biggr] S_1^2
        +\Biggl[
                -\frac{64 Q_{40}}{27 (N-1) N^2 (N+1)^2 (N+2)}
                \nonumber\\
                &&-128 S_2
        \Biggr] S_1^3
        +\frac{512}{9} S_1^4
        +\Biggl[
                \frac{4 Q_{175}}{27 (N-2) (N-1)^2 N^3 (N+1)^3 (N+2)^2}
                -384 S_3
        \Biggr] S_2
        \nonumber\\
        &&+\frac{32 \big(
                54+97 N+97 N^2\big) S_2^2}{3 N (N+1)}
        +\frac{32 (2+5 N) (3+5 N) S_4}{N (N+1)}
        +384 S_5
\nonumber\\ &&        
+\Biggl[
                \Biggl[
                        -\frac{256 \big(
                                2+3 N+3 N^2\big)}{N (N+1)}
+1024 S_1
                \Biggr] S_{-1}
                -256 S_{-2,1}
        \Biggr] S_{-2}
        +640 S_{-3} S_{-2}
        +192 S_{-2}^2
        \nonumber\\
        &&+
        \frac{64 \big(
                30+11 N+11 N^2\big) S_{3,1}}{3 N (N+1)}
        -384 S_{3,2}
        -384 S_{4,1}
        +\frac{256 \big(
                2+3 N+3 N^2\big) S_{-2,-1}}{N (N+1)}
        \nonumber\\
        &&-\frac{64 \big(
                6+11 N+11 N^2\big) S_{2,1,1}}{N (N+1)}
        +768 S_{3,1,1}
        -1024 S_{-2,1,-2}
\nonumber\\ &&        
+\frac{3 \big(
                2+N+N^2\big) 2^{8+N} S_{1,1}\big(\frac{1}{2},1\big)}{(N-2) N (N+1)^3 (N+2)}
-\frac{128 \big(
                2+N+N^2\big)^2 S_1(2) S_{1,1}\big(\frac{1}{2},1\big)}{(N-1) N^2 (N+1)^2 (N+2)}
        \nonumber\\
        &&
        -\frac{128 \big(
                2+N+N^2\big)^2 
(S_{1,2}\big(\frac{1}{2},2\big)-S_{1,1,1}\big(\frac{1}{2},1,2\big)-S_{1,1,1}\big(\frac{1}{2},2,1\big))}{(N-1) N^2 
(N+1)^2 (N+2)}
        \nonumber\\
&&
 +\Biggl[
                -\frac{32 S_1^2 Q_{12}}{3 (N-1) N (N+1)}
-\frac{4 Q_{174}}{27 (N-2) (N-1)^2 N^3 (N+1)^3 (N+2)^2}
                \nonumber\\
                &&
 +\ln(2) \Biggl[
                        -\frac{576 \big(
                                2+3 N+3 N^2\big)}{N (N+1)}
                        +2304 S_1
                \Biggr]
                        \nonumber\\
                        &&
                        +128 S_1^3
                +\Biggl[
                        \frac{16 Q_{133}}{27 (N-1) N^2 (N+1)^2 (N+2)^2}
                        +2432 S_2
                \Biggr] S_1
                -\frac{64 \big(
                        19+30 N+30 N^2\big) S_2}{N (N+1)}
                \nonumber\\
                &&+576 S_3
                +\Biggl[
                        -\frac{128 \big(
                                2+3 N+3 N^2\big)}{N (N+1)}
                        +512 S_1
                \Biggr] S_{-1}
                +192 S_{-2}
                +320 S_{-3}
                \nonumber\\
                &&-640 S_{-2,1}
                +\frac{64 \big(
                        2+N+N^2\big)^2 S_1(2)}{(N-1) N^2 (N+1)^2 (N+2)}
                +160 \zeta_3
        \Biggr] \zeta_2
                \nonumber\\
                &&
        +\Biggl[
                \frac{16 \big(
                        588+1171 N+1171 N^2\big)}{15 N (N+1)}
                        -\frac{6272}{5} S_1
        \Biggr] \zeta_2^2
        +\Biggl[
                \frac{64 S_1 Q_{34}}{3 (N-1) N (N+1)}
                +512 S_1^2
                \nonumber\\
                &&                -\frac{16 Q_{85}}{9 (N-1) N^2 (N+1)^2 (N+2)}
-192 S_2
                +192 S_{-2}
        \Biggr] \zeta_3
        -416 \zeta_5
\Biggr\},
\end{eqnarray}
%---------------------------------------------------------------------------------------------------------
with $Q_i$ polynomials in $N$, cf.~\cite{Ablinger:2020qvo}.

The numerical effect of the corrections of up to $O(\alpha^6 L^5)$ are illustrated in Table~\ref{TAB1} describing the 
$Z^0$-peak 
both in the case of a fixed width and the $s$ dependent width. The radiative corrections change both the peak 
positions and the width. 
The corrections due to the different orders may be positive or negative.
We list the individual contributions. 

At $O(\alpha^6 L^6)$ the corrections to the peak are 7\,keV and to the width 27\,keV. The next subleading 
correction serves as a control term. For measurements at the future FCC\_ee statistical
accuracies of $\Delta M_Z = 50~\keV~{\rm [stat]}, \Delta \Gamma_Z = 8~\keV~{\rm [stat]}$ 
are estimated \cite{DENT}. This precision level is about reached in the case of the present corrections.
In the case a further logarithmic order is needed, one needs to calculated both, one more order for the 
Drell--Yan process and one more for the massive OMEs. Furthermore, the electro--weak corrections and 
QCD corrections have to be pushed to higher orders. All of this will meet theoretical and technological 
(i.e. mathematical and computer--algebraic) challenges. 
%---------------------------------------------------------------------------------------------------------
\begin{table}[H]
    \centering
    \begin{tabular}{|l|r|r|r|r|}
    \hline
    \multicolumn{1}{|c|}{} &
    \multicolumn{2}{c|}{Fixed width} &
    \multicolumn{2}{c|}{$s$ dep. width} \\
    \hline
    \multicolumn{1}{|c|}{} &   
    \multicolumn{1}{c|}{Peak} &  
    \multicolumn{1}{c|}{Width}    &
    \multicolumn{1}{c|}{Peak} & 
    \multicolumn{1}{c|}{Width} \\  
    \multicolumn{1}{|c|}{} &   
    \multicolumn{1}{c|}{(MeV)} & 
    \multicolumn{1}{c|}{(MeV)}    &
    \multicolumn{1}{c|}{(MeV)} &
    \multicolumn{1}{c|}{(MeV)} \\
    \hline
    %lowest order                            &       &     &           &          \\
    $O(\alpha)$   correction                 &   185.638   &   539.408  &   181.098  &   524.978 \\
    $O(\alpha^2 L^2)$:                       & -- 96.894   & --177.147  & -- 95.342  & --176.235 \\
    $O(\alpha^2 L)$:                         &     6.982   &    22.695  &     6.841  &    21.896 \\
    $O(\alpha^2 )$:                          &     0.176   & --  2.218  &     0.174  & --  2.001 \\
    $O(\alpha^3 L^3)$:                       &     23.265  &    38.560  &    22.968  &    38.081 \\
    $O(\alpha^3 L^2)$:                       & --   1.507  & --  1.888  & --  1.491  & --  1.881 \\
    $O(\alpha^3 L)$:                         & --   0.152  &     0.105  & --  0.151  & --  0.084 \\
    $O(\alpha^4 L^4)$:                       & --   1.857  &     0.206  & --  1.858  &     0.146 \\
    $O(\alpha^4 L^3)$:                       &      0.131  & --  0.071  &     0.132  & --  0.065 \\
    $O(\alpha^4 L^2)$:                       &      0.048  & --  0.001  &     0.048  &     0.001 \\
    $O(\alpha^5 L^5)$:                       &      0.142  & --  0.218  &     0.144  & --  0.212 \\
    $O(\alpha^5 L^4)$:                       & --   0.000  &     0.020  & --  0.001  &     0.020 \\
    $O(\alpha^5 L^3)$:                       & --   0.008  &     0.009  & --  0.008  &     0.008 \\
    $O(\alpha^6 L^6)$:                       & --   0.007  &     0.027  & --  0.007  &     0.027 \\
    $O(\alpha^6 L^5)$:                       & --   0.001  &     0.000  & --  0.001  &     0.000 \\
    
    \hline
    \end{tabular}
    \caption[]{\sf Shifts in the $Z$-mass and the width due to the different contributions to the ISR QED
    radiative corrections for a fixed width of $\Gamma_Z =  2.4952~\GeV$  and $s$-dependent width using
    $M_Z = 91.1876~\GeV$
    \cite{PDG} and $s_0 = 4 m_\tau^2$, cf.~\cite{ALEPH:2005ab}; from \cite{Ablinger:2020qvo}.}
    \label{TAB1}
    \end{table}

The method of massive OMEs has been also applied to calculate the massive Wilson coefficients 
{for deep inelastic scattering}
in the region $Q^2 \gg m_Q^2$ in the single and two--mass cases at three--loop order analytically in Refs.~\cite{THRQCD}.  
%--------------------------------------------------------------------------------------------------------
\section{The Forward-Backward Asymmetry}
\label{sec:4}
%--------------------------------------------------------------------------------------------------------

\vspace*{1mm}
\noindent
A high luminosity measurement of the forward--backward asymmetry provides an excellent possibility
for a precision measurement of the fine structure constant at high energy \cite{Janot:2015gjr}. 
This is important because of its hadronic contributions \cite{Jegerlehner:2019lxt}.
The calculation of the forward--backward asymmetry at one--loop order has been performed in 
\cite{Bohm:1989pb,Beenakker:1989km,Bardin:1999gt,Bardin:1999ak,QED,Bardin:1989cw,Bardin:1990de,
Riemann:1989ca},
and the leading logarithmic two--loop corrections were computed in \cite{Beenakker:1989km}. More recently 
the leading logarithmic contributions to $O(\alpha^6 L^6)$ have been calculated in Ref.~\cite{Blumlein:2021jdl}.
Also the initial--final state interference terms, final state corrections, electro--weak and QCD corrections
are known at lower orders, cf.~\cite{Blumlein:2021jdl}.

The forward--backward asymmetry is defined as ratio over the difference and the sum of the angular 
integrals over the respective hemispheres measuring the final state muons,
%--------------------------------------------------------------------------------------------------------
\begin{eqnarray}
        \sigma_F = 2\pi \int\limits_{0}^{1} {\rm d} \cos(\theta) \frac{{\rm d}\sigma}{{\rm d} 
\Omega},~~~~~~ \sigma_B = 2\pi
        \int\limits_{-1}^{0} {\rm d} \cos(\theta) \frac{{\rm d}\sigma}{{\rm d} \Omega} \,.
\end{eqnarray}
%--------------------------------------------------------------------------------------------------------
The angle $\theta$ is defined between the incoming electron $e^-$ and the outgoing muon $\mu^-$ from 
$\gamma^*/Z^*$ decay, with
%--------------------------------------------------------------------------------------------------------
%--------------------------------------------------------------------------------------------------------
\begin{eqnarray}
        A_{\text{FB}}(s) &=& \frac{ \sigma_F(s) - \sigma_B(s)}{ \sigma_{T}(s) }, 
\end{eqnarray}  
%--------------------------------------------------------------------------------------------------------
and $\sigma_T(s) = \sigma_F(s) + \sigma_B(s)$. At Born level this reduces to \cite{BDJ}
%--------------------------------------------------------------------------------------------------------
\begin{eqnarray}
        \sigma_{\rm FB}^{(0)}(s) &=& \sigma_F^{(0)}(s) - \sigma_B^{(0)}(s) = \frac{\pi\alpha^2}{s} N_{C,f} 
\left(
        1-\frac{4m_f^2}{s} \right) G_3(s) \,,
\\
\sigma_{T}^{(0)}(s)  &=& \sigma_F^{(0)}(s) + \sigma_B^{(0)}(s) =
        \frac{4\pi\alpha^2}{3s} N_{C,f} \sqrt{1-\frac{4m_f^2}{s}} \left[ \left( 1 + \frac{2m_f^2}{s} 
\right) G_1(s)
- 6 \frac{m_f^2}{s} G_2(s) \right], 
\nonumber\\
\end{eqnarray}
%--------------------------------------------------------------------------------------------------------
with $m_f$ the final state fermion mass, $m_f \equiv m_\mu$ and  
%--------------------------------------------------------------------------------------------------------
\begin{eqnarray}
G_3(s) &=& 2 Q_e Q_f a_e a_f {\sf Re}[\chi_Z(s)] + 4 v_e v_f a_e a_f |\chi_Z(s)|^2.
\end{eqnarray}
%--------------------------------------------------------------------------------------------------------

The initial state corrections to $A_{\rm FB}$ {at leading logarithmic level} are described by two radiator functions 
$H_{e}^{LL}$ and 
$H_{FB}^{LL}$ \cite{Beenakker:1989km}, using the notation in \cite{Bohm:1989pb},
%--------------------------------------------------------------------------------------------------------
\begin{eqnarray}
        A_{\text{FB}}(s) &=& \frac{1}{\sigma_{T}(s)} \int\limits_{z_0}^{1} {\rm d}z \, \frac{4z}{(1+z)^2}
\tilde{H}_e^{LL}(z)
\sigma_{\rm FB}^{(0)}(z s),~~~~~~
        \sigma_{T}(s) = \int\limits_{z_0}^{1} {\rm d}z \, H_{e}(z) \sigma_{T}^{(0)}(z s) \,,
\end{eqnarray}
%--------------------------------------------------------------------------------------------------------
with
%--------------------------------------------------------------------------------------------------------
\begin{eqnarray}
\tilde{H}_e^{LL}(z) =
\left[ H_{e}^{LL}(z)
+ H_{FB}^{LL}(z)  \right],
\end{eqnarray}
%--------------------------------------------------------------------------------------------------------
where the parameter $z_0$ plays the role of an energy cut and $z = s'/s$. Here $H_{e}^{LL}$ is angular independent, and 
denotes the leading logarithmic contributions of the radiators of Section~\ref{sec:2}, while $H_{FB}^{LL}$ is 
angular dependent and obtained by the following integral
%--------------------------------------------------------------------------------------------------------
\begin{eqnarray}
        H_{FB}^{LL}(z) = \int\limits_{0}^{1} {\rm d} x_1 \int\limits_{0}^{1} {\rm d} x_2 \left( 
\frac{(1+z)^2}{(x_1+x_2)^2} -
        1 \right) \Gamma_{ee}^{LL}(x_2) \Gamma_{ee}^{LL}(x_1) \delta(x_1 x_2 - z). \label{eq:conv}
\end{eqnarray}
%--------------------------------------------------------------------------------------------------------
Here $\Gamma_{ee}^{LL}(x_i)$ denote the {massive operator matrix elements} and (\ref{eq:conv}) is expanded in 
$(\alpha L)$ to the desired order. It is convenient to consider first the Mellin--transform
%--------------------------------------------------------------------------------------------------------
\begin{eqnarray}
        \mathcal{M}[H_{FB}^{LL}(z)](n) &=& \int\limits_{0}^{1} {\rm d}z z^n H_{FB}^{LL}(z) = 
\int\limits_{0}^{1} {\rm d}x_1  
        \int\limits_{0}^{1} {\rm d}x_2 x_1^n x_2^n \left( \frac{(1+x_1x_2)^2}{(x_1+x_2)^2} - 1 \right) 
\Gamma_{ee}^{LL}(x_2)
        \Gamma_{ee}^{LL}(x_1)
         \nonumber \\
\end{eqnarray}
%--------------------------------------------------------------------------------------------------------
and to calculate the generating function 
%--------------------------------------------------------------------------------------------------------
\begin{eqnarray} \label{eq:GEN}
        \mathcal{G}[H_{FB}^{LL}(z)](t) &=& \sum\limits_{n=0}^{\infty} t^n \mathcal{M}[H_{FB}^{LL}(z)](n),
\end{eqnarray}
%--------------------------------------------------------------------------------------------------------
from which we then obtain the individual radiators. They can be expressed by harmonic and cyclotomic harmonic 
polylogarithms \cite{Ablinger:2011te} over the alphabet
%-------------------------------------------------------------------------------------------------------
\begin{eqnarray}
\label{eq:ALPH}
\mathfrak{A} = \left\{f_0 = \frac{1}{z}, f_1 = \frac{1}{1-z}, f_{-1} = \frac{1}{1+z},
f_{\{4,0\}} = \frac{1}{1+z^2}, f_{\{4,1\}} = \frac{z}{1+z^2} \right\}.
\end{eqnarray}
%-------------------------------------------------------------------------------------------------------
$H_{FB}^{LL}$ starts at $O(\alpha^2 L^2)$. 
%----------------------------------------------------------------------------
\begin{table}[H]\centering
        \begin{tabular}{rrrr}
                \toprule
                                                & $A_{\rm FB}(s_-)$         & $A_{\rm FB}(M_Z^2)$ & 
$A_{\rm FB}(s_+)$
                \\ \midrule
                $\mathcal{O}(\alpha^0)$         & $-0.3564803$          & $ 0.0225199$  & $0.2052045$
                \\
                $\mathcal{O}(\alpha L^1)$       & $-0.2945381$          & $-0.0094232$  & $0.1579347$
                \\
                $\mathcal{O}(\alpha L^0)$       & $-0.2994478$          & $-0.0079610$  & $0.1611962$
                \\
                $\mathcal{O}(\alpha^2 L^2)$     & $-0.3088363$          & $ 0.0014514$  & $0.1616887$
                \\
                $\mathcal{O}(\alpha^3 L^3)$     & $-0.3080578$          & $ 0.0000198$  & $0.1627252$
                \\
                $\mathcal{O}(\alpha^4 L^4)$     & $-0.3080976$          & $ 0.0001587$  & $0.1625835$
                \\
                $\mathcal{O}(\alpha^5 L^5)$     & $-0.3080960$          & $ 0.0001495$  & $0.1625911$
                \\   
                $\mathcal{O}(\alpha^6 L^6)$     & $-0.3080960$          & $ 0.0001499$  & $0.1625911$
                \\ \bottomrule
        \end{tabular}
        \caption{\sf $A_{\rm FB}$ evaluated at $s_-=(87.9\,{\rm GeV})^2$, $M_Z^2$
                        and $s_+=(94.3\,{\rm GeV})^2$ for the cut $z>4m_\tau^2/s$ from 
        Ref.~\cite{Blumlein:2021jdl}.}
        \label{tab:AFB1}; 
\end{table}
%----------------------------------------------------------------------------

\vspace*{-1cm}
\noindent
As an example $H_{FB}^{(3),LL}$  reads
%-------------------------------------------------------------------------------------------------------
\begin{eqnarray}
H_{FB}^{(3),LL}(z) &=& -\frac{16 (1-z) \big(4+11 z+4
 z^2\big)}{3 z} -\pi \biggl[
         \frac{4\big(2-3 z-2 z^2-3 z^3+2 z^4\big)}{3 z^{3/2}} \nonumber \\ && +\frac{4 (1-z) (1+5 
z)}{\sqrt{z}} H_0 +\frac{16
        (1-z)^2}{\sqrt{z}} H_{\{4,1\}} \biggr] +\biggl[
         \frac{4 (1+z) \big(5-18 z-19 z^2\big)}{3 z} \nonumber \\ && -\frac{16 (1-z) (1-7 z)}{\sqrt{z}} 
H_{\{4,0\}}
        -96 (1+z) H_{\{4,1\}} \biggr] H_0
-8 (1+z) H_0^2 \nonumber \\ && +\biggl[
         \frac{16 (1-z) (1+z)^2}{z} -\frac{64 (1-z)^2}{\sqrt{z}} H_{\{4,0\}} \biggr] H_1 +\biggl[ 
\frac{16\big(2-3 z-2 z^2-3
         z^3+2 z^4\big)}{3z^{3/2}}
\nonumber\\ &&
        +\frac{64 (1-z)^2}{\sqrt{z}} H_{\{4,1\}} \biggr] H_{\{4,0\}} +\biggl[ -\frac{16 (1-z) (1+z)^2}{z}
        +64 (1+z) H_0 \nonumber\\ && +\frac{64 (1-z)^2}{\sqrt{z}} H_{\{4,0\}} \biggr] H_{-1}
-64 (1+z) H_{0,1}
+ \frac{32 (1-z)^2}{\sqrt{z}} H_{0,\{4,0\}}
+96 (1+z) H_{0,\{4,1\}} \nonumber \\ &&
+ \frac{64 (1-z)^2}{\sqrt{z}} H_{1,\{4,0\}}
- \frac{64 (1-z)^2}{\sqrt{z}} H_{\{4,0\},\{4,1\}}
-64 (1+z) H_{-1,0} \nonumber \\ &&
- \frac{64 (1-z)^2}{\sqrt{z}} H_{-1,\{4,0\}}
+20 (1+z) \zeta_2,
\end{eqnarray}
 %-------------------------------------------------------------------------------------------------------
with $H_{\vec{w}} \equiv H_{\vec{w}}(\sqrt{z})$ and $\zeta_k, k \in \mathbb{N}, k \geq 2$ are the values 
of Riemann's $\zeta$ function at integer argument.

Let us give some numerical illustration. We compute $A_{\rm FB}(s)$ for different values around the 
$Z^0$-peak as suggested in \cite{Janot:2015gjr} to higher orders and see the corresponding improvement in 
Table~\ref{tab:AFB1}.
Comparing the results at one--loop to the highest order we obtain corrections of $-3$ \% for $s_-$ and  $-1 \%$ for 
$s_+$. More numerical studies are presented in Ref.~\cite{Blumlein:2021jdl}.
%--------------------------------------------------------------------------------------------------------
\section{Conclusions}
\label{sec:5}
%--------------------------------------------------------------------------------------------------------

\vspace*{1mm}
\noindent
The precision calculations of the QED initial state corrections to the process $e^+e^- \rightarrow
\gamma^*/Z^*$ are an essential ingredient for the planned key measurements at high energy $e^+e^-$ 
colliders such as the ILC, CLIC, FCC\_ee, and future muon colliders. Due to the smallness of the 
ratio $m_l^2/s$, with $m_l$ the initial charged lepton mass, the logarithmic and constant term corrections in a 
massive environment are 
sufficient. It therefore has been important to clarify the differences between 
Ref.~\cite{Berends:1987ab} and Ref.~\cite{Blumlein:2011mi} at $O(a^2)$. The differences found 
do in principle require to repeat the LEP electro--weak analysis, given the current accuracy of
the $Z^0$ peak and width, because of the respective theoretical shifts.
Since the codes {\tt TOPAZ0} \cite{Montagna:1998kp,Bardin:1999ak} and {\tt ZFITTER} \cite{ZFITTER}
contain the results of \cite{Berends:1987ab}, they have to be updated for the use in further experimental 
analyses.

The agreement between Refs.~\cite{Blumlein:2020jrf} and \cite{Blumlein:2011mi} allowed to use 
the method of massive OMEs for even higher order corrections in the fine structure constant. Having 
available all OMEs which contribute  {at two--loop order}, 
the first three {logarithmic} expansion coefficients
are available through the renormalization group equations to any order in the fine structure 
constant, i.e. two further orders beyond the leading logarithmic approximation. The calculation of 
one more order seems to be technically possible. The presently available corrections reach the 
projected accuracies at the FCC\_ee and need to be supplemented by corresponding QCD and electro--weak 
corrections. 

For the forward--backward asymmetry, which might allow a precision measurement of the fine structure 
constant in the future, the leading logarithmic corrections have been extended to $O((\alpha L)^6)$ and one
might want to consider sub--leading corrections. Beyond the corrections in the leading logarithmic 
approximation, where harmonic sums are sufficient to express the radiators in Mellin space, one also 
finds generalized and cyclotomic harmonic sums, forming a part of the function spaces having been
already revealed in other analytic massive single--scale calculations \cite{Blumlein:2018cms,THRQCD}.
We are still in the early phase to calculate the necessary radiative corrections for the 
high--luminosity measurements at the FCC\_ee and will view more theoretical results during the 
decades ahead.

\vspace*{5mm}
\noindent 
{\bf Acknowledgments.}\\
We would like to thank A.~De Freitas,  W.L.~van Neerven, and C.G.~Raab for collaboration on the present 
topic and P.~Janot and G.~Passarino for discussions. This project has received funding from the European 
Union's Horizon 2020 research and innovation programme under the Marie Sk\/{l}odowska-Curie grant agreement 
No. 764850, SAGEX.
%-------------------------------------------------------------------------------------

%-----------------------------------------------------------------------------------
\end{document}